\documentclass[acmsmall,screen, nonacm]{acmart}

\AtBeginDocument{%
  }

\setcopyright{none}
\copyrightyear{2026}
\acmYear{2026}
\acmDOI{XXXXXXX.XXXXXXX}
\acmISBN{}
\acmPrice{}
\acmJournal{CSUR}

\usepackage[T1]{fontenc}

\usepackage[capitalize]{cleveref}
\crefname{section}{Sec.}{sections}
\Crefname{section}{Section}{Sections}
\usepackage{xcolor}
\usepackage{relsize}
\usepackage{booktabs}
\usepackage{hyperref}
\usepackage{microtype}
\usepackage{framed}
\usepackage{tikz}
\usetikzlibrary{backgrounds}

\usepackage{pgfplots} 
\usepackage{pgfplotstable}
\usepackage{sansmath} 
\pgfplotsset{compat=1.18}

\usepackage{tabularx}

\usepackage[most]{tcolorbox}
\tcbset{
  myexample/.style={
    enhanced,
    breakable,               
    colback=white,
    colframe=black,
    sharp corners,
    boxrule=.5pt,
    fonttitle=\itshape\color{black},
    title={\textit{Running Example}},
    before skip=1pt,
    attach boxed title to top left={xshift=5mm,yshift=-3mm},
    boxed title style={
      colback=white,
      colframe=white,
      top=1pt,
      bottom=0pt,
      left=0pt,
      right=0pt, 
      boxsep=0pt
    },
    boxsep=2pt,
    left=4pt,
    right=4pt,
    bottom=2pt,
  }
}

\usepackage[textsize=scriptsize, backgroundcolor=red!50, linecolor=red!50]{todonotes}

\newcommand\parhead[1]{\vspace{+.2mm}\noindent\textbf{{#1}}.}
\newcommand{\code}[1]{\texttt{\small #1}}

\begin{document}


\title{Before Autonomy Takes Control: Software Testing in Robotics}

\author{Nils Chur}
\orcid{0009-0009-2427-8342}
\email{nils.chur@rub.de}
\affiliation{%
	\institution{Ruhr University Bochum}
	\city{Bochum}
	\country{Germany}
}

\author{Thiago Santos de Moura}
\orcid{0009-0007-1927-8378}
\email{thiago.santosdemoura@rub.de}
\affiliation{%
	\institution{Ruhr University Bochum}
	\city{Bochum}
	\country{Germany}
}

\author{Argentina Ortega}
\orcid{0000-0002-3873-4435}
\email{argentina.ortega@uni-bremen.de}
\affiliation{%
	\institution{University of Bremen}
	\city{Bremen}
	\country{Germany}
}
\affiliation{%
  \institution{Ruhr University Bochum}
  \city{Bochum}
  \country{Germany}}

\author{Sven Peldszus}
\orcid{0000-0002-2604-0487}
\email{sven.peldszus@rub.de}
\affiliation{%
	\institution{Ruhr University Bochum}
	\city{Bochum}
	\country{Germany}
}
\affiliation{%
	\institution{IT University of Copenhagen}
	\city{Copenhagen}
	\country{Denmark}
}

\author{Thorsten Berger}
\orcid{0000-0002-3870-5167}
\email{thorsten.berger@rub.de}
\affiliation{%
  \institution{Ruhr University Bochum}
  \city{Bochum}
  \country{Germany}
	}
 \affiliation{%
 	\institution{Chalmers\,|\,Gothenburg University}
 	\city{Gothenburg}
 	\country{Sweden}}

\author{Nico Hochgeschwender}
\orcid{0000-0003-1306-7880}
\email{nico.hochgeschwender@uni-bremen.de}
\affiliation{%
  \institution{University of Bremen}
  \city{Bremen}
  \country{Germany}
}

\author{Yannic Noller}
\orcid{0000-0002-9318-8027}
\email{yannic.noller@rub.de}
\affiliation{%
	\institution{Ruhr University Bochum}
	\city{Bochum}
	\country{Germany}
}

\renewcommand{\shortauthors}{Chur et al.}


\begin{abstract}
    Robotic systems are complex and safety-critical software systems. As such, they need to be tested thoroughly. Unfortunately, robot software is intrinsically hard to test compared to traditional software, mainly since the software
    needs to closely interact with hardware, account for uncertainty in its operational environment, handle disturbances, and act highly autonomously.
    However, given the large space in which robots operate, anticipating possible failures when designing tests is challenging.
    This paper presents a mapping study by considering robotics testing papers and relating them to the software testing theory.
    We consider 247 robotics testing papers and map them to software testing, discussing the state-of-the-art software testing in robotics with an illustrated example, and discuss current challenges.
    Forming the basis to introduce both the robotics and software engineering communities to software testing challenges.
    Finally, we identify open questions and lessons learned.

\end{abstract}



\maketitle



\section{Introduction}%
\label{sec:introduction}
Robotic systems are on the rise, operating across diverse domains such as healthcare, agriculture, and transportation, making reliability and safety crucial.
These systems combine hardware and software components to act autonomously in dynamic and safety-critical environments\,\cite{garcia2020esec}. 
Consequently, thorough testing before deployment is required to uncover failures---``\emph{the inability of the system to perform its required functions within specified performance requirements}''\,\cite{ieee61012}---but also to evaluate performance, standards conformance, and ensuring their trustworthiness and reliability.
Inadequate testing can lead to personal injury, damage to infrastructure, and loss of profit.

\looseness=-1
Testing robotic systems poses unique challenges, as the software has to handle sensor noise, actuator variability, and other external influences
that introduce non-determinism, which may lead to incorrect behavior or unexpected failures\,\cite{afzal2020icst}.
In addition to these robotics-specific challenges\,\cite{Brugali2006Stability,sven2025reconfiguration,garcia.ea:2023:robotvar,garcia2019robotics}, robotics software faces the same challenges as traditional software systems, requiring various components to work seamlessly together under uncertain and changing conditions.
Consequently, validating the behavior of robotic systems across a variety of environmental and operational conditions remains challenging. 
Testing in robotics is seen as fragmented and largely generic, often ad hoc\,\cite{araujo2022tse}, whereas software engineering has developed comprehensive testing standards\,\cite{iso29119pt1,iso29119pt2,iso29119pt3,iso29119pt4}, whose adaptation in robotics remains unclear.
%
To the best of our knowledge, no systematic overview exists that relates robotics testing to software testing standards.
Current literature focuses on specific applications, case studies, or single testing methods, but lacks a comprehensive overview of the testing process or appropriate techniques.  
Although recent studies examined software engineering practices in robotics\,\cite{garcia2020esec,brugali2007software,educon2014SPL} to better understand the particular challenges of robotics testing\,\cite{afzal2020icst,song2021wain,afzal2021icst,araujo2022tse}, little work provides a concise introduction to robotics testing.

We address this gap through a systematic literature review of robotics testing, analyzing which testing techniques and standards from software engineering have been applied and how.
Our goal is to introduce established software testing theory to the robotics community while presenting the unique characteristics of robotics testing to the software engineering community, thereby identifying gaps and challenges.

\section{Background and Motivation}\label{sec:background-motivation}
\looseness=-1


Robotic systems comprise tightly coupled software and hardware components, such as planning and control modules, sensors, and actuators.
Their interdisciplinary nature, spanning computer science, control theory, and mechanical and electrical engineering, complicates testing, which must account for heterogeneous components and their intricate interactions, creating numerous potential failure points.
This challenge is compounded by the frequent absence of formal software engineering practices in robotic development, limiting the application of rigorous testing\,\cite{garcia2020esec}.

\looseness=-1
Inadequate testing increases the likelihood of failures, the inability of a software system to perform its intended functions, such as producing incorrect outputs. 
Failures arise from underlying faults, such as erroneous or missing code\,\cite{ieee61012}, and are identified through testing. 
%
In robotics, failures often emerge from interactions among software components, individual component faults, or unpredictable environments. 
The use of middleware, while facilitating component integration, introduces additional complexity, making root-cause analysis essential for isolating the precise sources of failure.
\looseness=-1
Unlike traditional software failures, which might cause crashes or incorrect outputs, robotic software failures can cause physical harm or property damage.
Operating alongside humans in domains such as healthcare, or autonomous driving makes these systems safety-critical\,\cite{bozhinoski2019jjss}, requiring rigorous safety validation, often in compliance with strict industry standards\,\cite{bi2020rcim}.

\subsection{Testing}
\looseness=-1
\textbf{Software testing} can be divided into \textit{static testing}, which examines the software without executing it, and \textit{dynamic testing}, which involves running the software to detect failures. 
This paper focuses on dynamic testing, defined by \citet{lochau2014sfm} as the “dynamic validation/verification of the behavior of a program on a finite set of test cases suitably selected from the usually infinite input/execution domain against the expected behavior.”
In this context, test cases are discrete, repeatable units that enable systematic evaluation of software behavior\,\cite{iso29119pt1}.
Each test case consists of four elements:

\looseness=-1
\textit{(I) Precondition}: The state before test execution, including a robot's environment, data, software, and hardware~\cite{iso29119pt1}. 
This comprises a robot’s initial pose, pose-estimation belief, map, system configuration, and test environment. 
We distinguish between environment representation, i.e., inputs modeling a robot’s surroundings, and the test environment, which includes all resources required to execute a family of tests, such as simulators, hardware settings, and middleware configurations.
    
\textit{(II) Test Input}: Data required for test execution, such as commands or stimuli applied to the system, ranging from simple function parameters to whole simulation environments.

\textit{(III) Expected Outcome}: The expected system behavior or state during test execution, used to determine whether the test objective has been met.
    
\textit{(IV) Postcondition}: The state of the system and environment after test completion, including ``success criteria, failures to check for, etc.''\,\cite{iso29119pt1}, such as confirmation signals, logs showing correct behavior, or error messages indicating test failures.

\textbf{Testing robotic systems} remains challenging, as they operate in highly variable and continuous environments, unlike other cyber-physical systems deployed in more controlled settings.
This complicates to systematically derive corner cases\,\cite{afzal2020icst} and specify expected behaviors\,\cite{aliabadi2020ikt}. 
Although testing of cyber-physical systems has been studied\,\cite{garousi2018testing}, practical guidance tailored to the specific characteristics of robotic systems remains limited---particularly for practitioners without a background in software testing or software engineers lacking robotics expertise.
This paper aims to bridge that gap by providing an overview of the key concepts, test levels, types, and activities relevant to robotics testing, supported by concrete examples.

\subsection{A Case Study: Restaurant Robots}\label{sec:example}
\looseness=-1
For illustration, our running example is a service robot operating in a restaurant. 
It is inspired by challenges in the RoboCup@Home competition, 
which aims to advance the development of personal service robots and benchmarking their performance in realistic home environments.
Additional inspiration comes from commercial robots, such as Bella 
and Servi. 

\noindent
\begin{minipage}[t]{0.57\textwidth}
\vspace{0pt}
\begin{tcolorbox}[
  colback=gray!10,
  colframe=gray!50,
  boxrule=0.5pt,
  arc=2mm,
  left=4pt,
  right=4pt,
  top=4pt,
  bottom=4pt
]
\textbf{Restaurant Robot:}
The robot assists waiters by collaboratively taking and delivering orders and cleaning tables. 
It is a PAL Robotics TIAGo (\cref{fig:tiago}) with a differential-drive base, laser sensors for localization and obstacle detection, a 7-degree-of-freedom arm with a parallel gripper, an onboard computer, an RGB-D camera for perception, and a speaker, microphone array, and touchscreen for user~interaction.
\end{tcolorbox}
\end{minipage}
\hfill
\begin{minipage}[t]{0.4\textwidth}
\vspace{0pt}
\centering
\includegraphics[width=\linewidth]{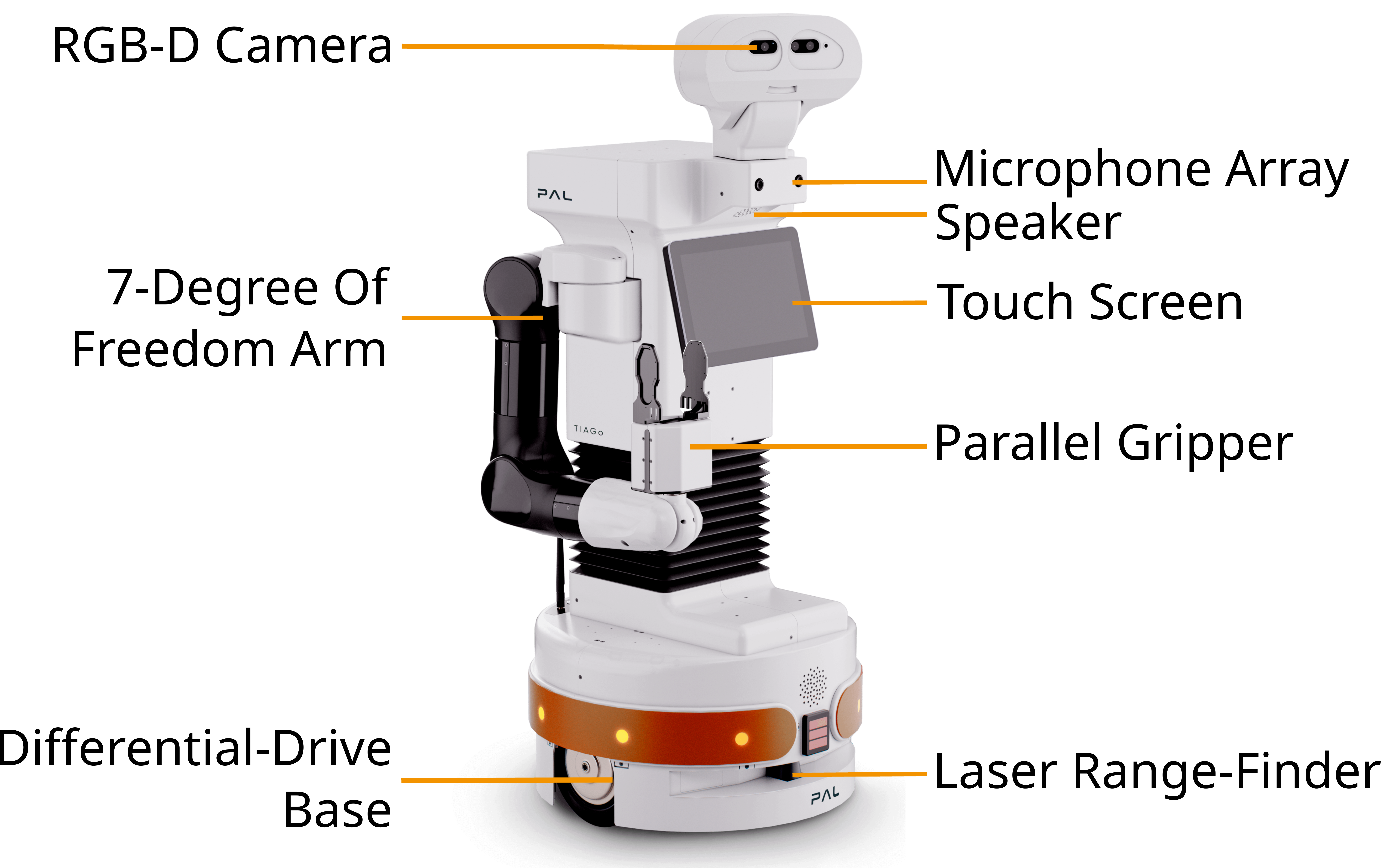}
\vspace{-22pt}
\captionof{figure}{PAL Robotics TIAGo robot\,\cite{macenski2023regulated}\label{fig:tiago}}
\end{minipage}
\vspace{4pt}

\parhead{Tasks}\label{sub:sec:application}
We focus on three tasks, in which the robot must navigate freely within a restaurant while avoiding collisions with furniture and people. 
Its velocity must remain safe and adapt when transporting food or liquids. 
The robot should monitor its battery and recharge to remain available. 
\Cref{tab:requirements} presents an (incomplete) list of mission requirements needed to realize the three tasks.



\begin{table}[t]
  \centering
  \scriptsize
  \caption{Requirements for our running example}\label{tab:requirements}
  \setlength{\tabcolsep}{2pt}
	\vspace{-.3cm}
\begin{tabularx}{\textwidth}{rp{0.34\textwidth}X}
  \toprule
  \textsf{ID} & \textsf{feature} & \textsf{requirement} \\
  \midrule
  1 & Receive food order from app/kitchen staff & The robot shall add the new ``pick up and delivery'' task to its task queue.\\
  2 & Receive request over the network to go to a table & The robot shall add the ``take order'' task to its task queue with a high priority. \\
  3 & Recognize tables with people ready to order & The robot shall recognize hand gestures from people at the table.\\
  4 & Take order through on-robot touch screen & The robot shall communicate that it has received the order.\\
  5 & Takes order verbally & The robot shall listen to the order and create a machine-readable food order task.\\
  6 & Pick up orders from kitchen & The robot shall wait for confirmation that the food order has been correctly loaded.\\
  7 & Deliver order to table & The robot shall detect that food orders have been unloaded.\\
  & & The robot should ask for confirmation that food orders are correctly delivered.\\
  8 & Pick cutlery and dishes & The robot shall autonomously collect all dishes and cutlery left at an empty table. \\
  & & The robot shall put dirty dishes and cutlery in the sink.\\
  9 & Clean surfaces & The robot shall move its arm following a grid path while holding a sponge.\\
  10 & Navigation  & The robot shall be able to autonomously reach a subset of pre-defined locations (e.g.~customer tables, kitchen pickup and drop-off locations). \\
  & & The robot shall not crash into people.\\
  & & The robot shall not exceed the maximum safety velocity of 0.8 m/s.\\
  11 & Recharge battery & Throughout the entire operation, the robot should add ``battery recharge'' tasks to its task queue to maintain at least 40\% of its total battery capacity. \\
  \bottomrule
\end{tabularx}
\vspace{-.5cm}
\end{table}

\textit{1. Getting customer orders.}
The robot receives customer orders via an app, or customers summon it via an app or hand gestures. Orders can be entered through the robot’s tablet or using speech recognition.
Kitchen staff can request the robot to pick up and deliver orders to specific tables. 

\looseness=-1
\textit{2. Pick up and delivery of orders.}
The robot navigates from its current location to a predefined pickup point, where staff loads the order. 
The robot waits for confirmation that the order has been successfully loaded before proceeding. 
Once confirmed, it navigates from the kitchen to the designated table and waits for staff or customers to unload the order.

\textit{3. Clean up tables.} \looseness=-1 
The robot picks up leftover dishes and cutlery, places them on its tray, and brings them back to the kitchen.
It can also clean tables using a sponge and dry them with a dish towel.

\parhead{Software}\label{software}
\looseness=-1
Our robot uses the Robot Operating System (ROS)\,\cite{quigley2009icra}, a middleware and robot application framework. A ROS application consists of so-called nodes (i.e., independent software components), which communicate using publish-subscribe and other communication mechanisms.
For instance, a LaserNode publishes sensor readings to a so-called ROS topic, and a LocalizationNode subscribes to it and performs computation when receiving new data.

\begin{figure}[t]
\vspace{-.3cm}
  \includegraphics[width=\textwidth]{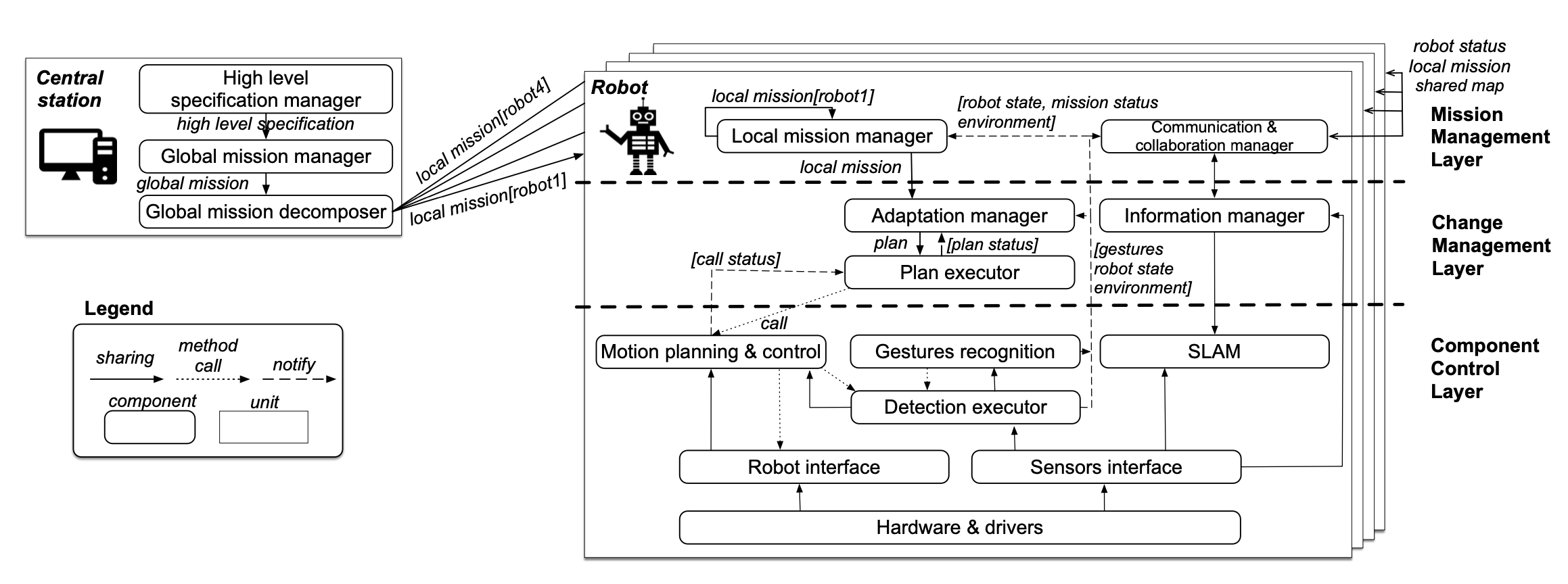}
	\vspace{-.8cm}
  \caption{A self-adaptive decentralized robotic architecture (SERA)\,\cite{garcia2018icsa}\label{fig:sera-architecture}}
  \Description{A diagram of the SERA architecture that shows the components at the three levels. Highest level deals with mission and communication. Next the middle layer contains adaptation and information managers, and the plan executor. Finally, the lowest layer includes the components used in the execution, e.g.~motion planning, SLAM, and hardware and drivers.}
	\vspace{-.6cm}
\end{figure}

\looseness=-1
For our example robot, we use the self-adaptive decentralized robotic architecture (SERA)\,\cite{garcia2018icsa}, a three-layer architecture comprising mission management, change management, and component control (\cref{fig:sera-architecture}).
At the mission management layer, service goals, such as delivering an order to a specific table, are processed and decomposed into actionable tasks, which are coordinated by the local mission manager.
The change management layer monitors mission execution and reacts to environmental or operational changes, such as blocked paths or human interaction. 
The adaptation manager evaluates these situations and triggers replanning or rescheduling as needed, while the plan executor dispatches tasks to the underlying control components.
The component control layer includes navigation, motion control, and perception components, and interfaces with hardware and sensors.


\looseness=-1
Each component comprises multiple nodes that interacted with other nodes across layers, forming systems of systems.
For example, the LocalizationNode process laser scans from the LaserNode but also odometry data, maintains a probabilistic belief over the robot's pose, and continuously publishes pose estimates that are used by navigation for obstacle avoidance.
Similarly, the motion planning \& control component simultaneously subscribes to localization updates, map data, and obstacle information, while publishing control commands and status feedback to the plan executor component.
Different update rates between nodes and independent and concurrent execution introduce further complexities into the message exchange.
Therefore, the behavior of a single component is influenced by other components, the current mission, and the adaptation context.
This tight coupling through asynchronous communication, combined with parallel execution across layers, and often required real-time capabilities, shows why robotics software is complex.

\section{Methodology}%
\label{sec:methodology}
\looseness=-1
Our methodology combines expert knowledge with a survey of the literature to review the state-of-the-art in robotics testing.
In particular, we compared the state of the art in software testing, as captured in ISO testing standards\,\cite{iso29119pt1,iso29119pt2,iso29119pt3,iso29119pt4}, with testing practices in the robotics testing literature. Then, based on our expert knowledge, we characterized the overlap.
To this end, we brought together experts from multiple relevant fields.
The resulting authors comprised experts in robotics, testing in general, and software engineering, with most having expertise in multiple of these areas.
This expertise allowed us to not only provide a comprehensive overview, but also put the existing literature and the practices they knew from experience into context.
Among others, this complementary expertise allowed us to identify gaps in the current state of the art in robotics software testing.

\looseness=-1
We used the body of knowledge on testing from the ISO\,29119 standard family (mainly parts 1--4)\,\cite{iso29119pt1,iso29119pt2,iso29119pt3,iso29119pt4} to systematically identify the aspects of testing considered relevant by the community.
Based on our expertise in testing and additional literature, e.g., the documentation of the ISTQB Certified Tester\,\cite{istqb2018,istqb-glossary} and academic testing papers or books\,\cite{lochau2014sfm, hass2014guide, bath2014}, all authors discussed at which level of abstraction to consider testing aspects and decided, for instance, to merge all low-level coverage criteria introduced in the ISO standard. 

\begin{figure}[b]
\vspace{-.5cm}
    \centering
    \includegraphics[clip, trim=0.9cm 0.1cm 0.7cm 0.1cm, width=\linewidth]{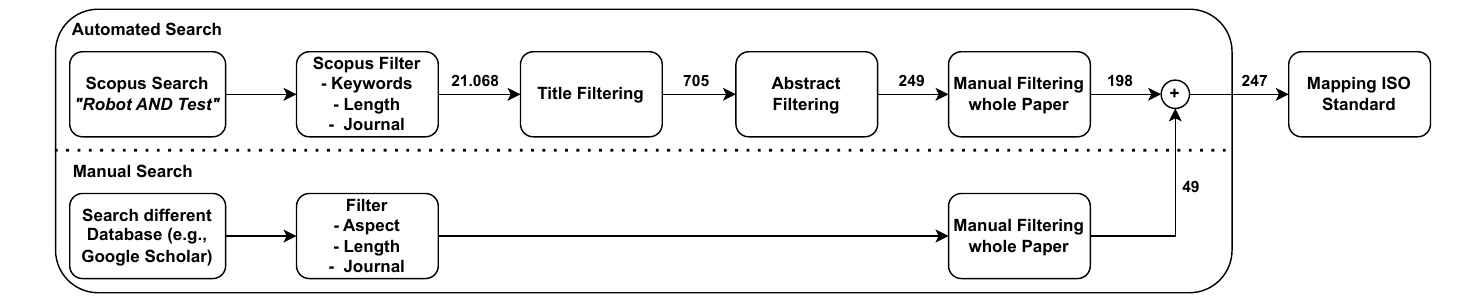}
		\vspace{-.7cm}
    \caption{Overview of our search strategy and selection process}
    \label{fig:selectionProcess}
    \Description{Figure 3. Fully described in the text.}
		\vspace{-.4cm}
\end{figure} 

\looseness=-1
\Cref{fig:selectionProcess} shows our strategy. We searched Scopus for papers focusing on testing robotic systems, so those that introduce or evaluate testing tools, propose testing techniques, or report on case studies that explicitly address verification and validation.
We further included studies that discuss challenges, limitations, or open problems related to testing.
To capture all possible dimensions in respect to testing and robotics, we identified papers using the general search string \textit{``Robot AND Test''} on paper titles and abstracts.
We limited the search to the categories `computer science' and `engineering' and required that the papers have any robotic-related author keywords, such as navigation, manipulation, or path planning---selected by skimming the full list of 161 keywords in Scopus and added to our search term.
To ensure quality, we only considered full papers with at least six pages, published in journals or conference proceedings.
This search resulted in 21,068 papers potentially reporting about testing of robots or testing techniques for robots.
To identify the papers that in fact report about testing robots, we applied a two-step filtering, first based on the title, and then on the abstract.
We filtered 60 papers independently by two authors each, resulting in agreement for 50\% of them on whether to exclude or include the paper. 
Disagreements were discussed among all authors.
Due to the low level of agreement and little information contained in the titles, we decided to only exclude obvious cases, resulting in 705 papers after the initial filtering based on titles.
Next, we applied a second round of filtering based on the abstract.
This time, 160 papers were filtered by two authors each, this time achieving an agreement in 71\% of the cases. 
Again, disagreements were discussed among all authors before filtering the papers independently thereafter, resulting in 249 included papers after abstract-based filtering.

\looseness=-1
We mapped the testing-related concepts of the testing community to the papers identified, thereby obtaining an overview of what testing aspects are covered by the robotics community.
Thereby, we excluded 17 papers because we could not access them, 12 because they were below our size limit (which was not in the meta-data), and 28 papers because they did not describe a relevant aspect of robot testing related to the testing body of knowledge, despite the abstract indicating so. 
This resulted in 198 papers on testing in robotics, of which 12 report on testing challenges, and the remaining 186 we mapped to the testing body of knowledge.
Additionally, we conducted a manual search for all aspects of the testing body of knowledge with Google Scholar for robotics papers based on our expertise, especially for aspects that were not covered by the initial search, leading to 49 additional papers.
To broaden the scope of this search, we intentionally did not use Scopus again.

\looseness=-1
Based on the mappings between testing body of knowledge and the robotics literature, details from the papers, and our experiences in robotics, we then summarized the state of the art of testing in the robotics domain.
Thereby, we identified testing aspects not covered by the robotics community, discrepancies in the understanding of specific testing aspects, and aspects covered but not as systematically as intended in the testing community.
These observations build the basis for an informed discussion, particularly leveraging the different perspectives resulting from the backgrounds of the authors, to derive recommendations on how to improve the testing of robotic systems.

\section{Test Environment, Levels, Types, and Activities}

Testing comprises a set of activities that together form the test process\,\cite{bertolino2007software}, deciding what should be tested, how, and in which environment.
We follow a typical classification of tests introduced by software testing standards\,\cite{iso29119pt1,iso29119pt2,iso29119pt3,iso29119pt4}.
First, we introduce the test environment in \cref{sec:sub:test-envi}, that is, context in which tests are executed. 
Thereafter, test levels are introduced in \cref{sec:sub:test-levels}, representing different stages of testing that align with the system's development process.
Each test level focuses on different parts of the system, from single units to the whole system.
Then, test types are presented in \cref{sec:sub:test-types}, describing the purpose of the test and whether it accomplishes the intended functionality or quality under various conditions.
Finally, test activities will be discussed in \cref{sec:sub:test-activities}, providing a systematic test process, from planning and implementing to execution and evaluation. 

\begin{figure}[b]
\vspace{-.6cm}
  \includegraphics[clip, trim=0.5cm 8.5cm 0.5cm 0.5cm, width=\textwidth]{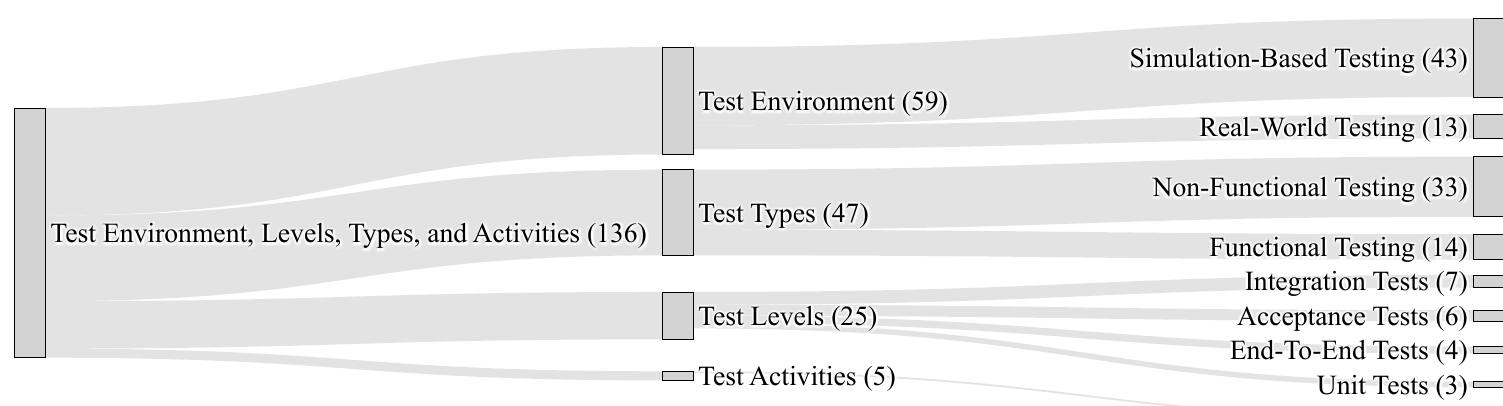}
	\vspace{-.5cm}
  \caption{Number of robotics papers reporting about the respective testing aspect}
  \Description{A Sankey diagram that visualizes the classification of papers discussed in this survey.}%
  \label{fig:subsectionPapers-Part1}
	\vspace{-.5cm}
\end{figure}

\looseness=-1
We found, see \cref{fig:subsectionPapers-Part1}, that the majority of papers focus on the test environment (59), of which most are simulation-based testing (43) followed by real-world testing (13). 
Under test types, most papers discuss non-functional testing (33) with a focus on non-functional performance testing (15); or they discuss functional testing (14).
Research on test levels (26), divided into integration tests (7), acceptance tests (6), end-to-end tests (4), and unit tests (3), as well as testing activities (5) is sparse.

\subsection{Test Environment}\label{sec:sub:test-envi}
\looseness=-1
The environment in which tests are executed is a particular aspect of testing robots compared to traditional software systems.
It can either be a simulation-based environment that mimics the operational environment, or the real one.
Choosing a test environment impacts various aspects.
While simulation-based environments offer advantages in terms of cost and speed, they do not provide the same level of realism as testing in a real environment\,\cite{afzal2021icst, sotiropoulos2017qrs, neri2023experimental, rodic2011scalable}. 


\parhead{Simulation-Based Testing}\label{sub:sec:sim-based}
\looseness=-1
Simulation-based testing refers to executing tests within a virtual environment, which can range from simple mocks to high-fidelity simulations.
The developed software is executed on a virtual model of the robot, including sensors and actuators, and its dynamic environment.
Simulation-based testing is widely used in robotics due to its scalability, repeatability, and cost-efficiency, especially for domains where real-world tests are impossible to conduct\,\cite{afzal2021icst, sotiropoulos2017qrs,liaqat2019autonomous, parra2023iros}. It enables testing diverse scenarios without the risks associated with real-world testing, such as property damage, and allows for automated evaluation and regression testing\,\cite{teixeira2021cloud, jimenez2023automated, araiza2016systematic, sartori2022integration}.

\looseness=-1
Testing in virtual environments comes with the drawback of a limited realism, as it cannot fully replicate the complexity of the real world. Although traditional software testing often relies on virtual environments to simulate execution conditions, robotic simulations struggle to model real-world physics, sensor inaccuracies, and dynamic environmental conditions\,\cite{selecky2019analysis, robert2020irc}. Factors such as friction, material deformation, and sensor noise also add differences between the simulated and real-world performance, also known as the simulation-reality gap\,\cite{aljalbout2025reality}.

\looseness=-1
Research in this direction focuses on building simulation, addressing the simulation-reality gap by designing high-fidelity simulations for different robots, such as manipulators and humanoids\,\cite{sadka2020virtual, seyyedhasani2020collaboration, lee2022towards, platt2022comparative, wolf2020evolution}, ground vehicles\,\cite{sobczak2022building, shell2020reality}, or aerial vehicles\,\cite{maftei2020multi}.
Resulting in many frameworks simplifying simulation setup or robot integration\,\cite{douthwaite2021modular, munawar2019real, sibilska2022framework, wilson2012uav}, such as ground vehicles\,\cite{balakirsky2008integrated, schiegg2019novel, sotiropoulos2016virtual, vieira2019copadrive}, and space robots\,\cite{jochmann2014virtual}.
%
Simulations can also be used to evaluate the robotic system\,\cite{mihali2000robotic} or components, ranging from control\,\cite{insam2020high, marcosig2018devs, cheng2023simulation, chung2021hardware, moshayedi2021simulationPID, sartori2014implementation}, cooperation and formation\,\cite{mercier2020terrestrial, seyyedhasani2020collaborationPartII, neri2023experimental}, to perception\,\cite{moshayedi2021simulation} and navigation\,\cite{pinrath2020integration, megalingam2020comparison}.
Additionally, research is conducted to test the safety of the robot in human-robot interaction\,\cite{huck2020simulation, zenzeri2013using}, or multi-robot systems\,\cite{harbin2021model, dey2022synchrosim, rodic2011scalable}.
This only provides a small fraction of the complexity and a founded basis, however, grasping the full complexity of simulation-based testing is beyond the scope of our work.

\parhead{Real-World Testing}
By executing the software on actual robot hardware and letting it operate directly in the physical environment, real-world testing is essential for verifying real performance\,\cite{xu2021four, marquez2023hardware} and the correct behavior\,\cite{kim2023development, palermino2022development, mashrur2023assistive, icinco08}. Assumptions made in simulation, issues related to hardware integration, and environmental variability can be identified.
Research addresses developing benchmarks for systems for better comparison between solutions\,\cite{vitzilaios2009experimental, borenstein1995umbmark}, and safety in human-robot interaction\,\cite{punzo2019bipartite, torta2012modeling}.
Real-world testing is often used for higher test levels, since the complete system needs to be available.
The actual test environment does not necessarily have to be the one the robot was designed for and can be a laboratory or a controlled real-world environment.
Deploying the robot in the environment it is designed for is a special case of real-world testing, typically referred to as field-based testing, discussed in \cref{sec:sub:field-based-testing}.

\looseness=-1
Real-world testing offers the highest fidelity and realism, but it is costly, less reproducible, and time-consuming\,\cite{shin2022hands}. A combination of both simulation-based and real-world testing is employed to leverage both fidelity and cost\,\cite{zofka2020pushing}. Simulation-based testing during development and real-world testing for validation\,\cite{afzal2021icst, afzal2020icst, mannion2020introducing}. In the remainder we focus on different testing aspects, whereby the test environment plays a secondary role and will therefore not be discussed explicitly. 

\subsection{Test Levels}
\label{sec:sub:test-levels}
A test level classifies testing activities by the scope and granularity of the system under test, specifying where testing is applied and how objectives are aligned with that scope.

\begin{figure}
	\includegraphics[width=.8\textwidth]{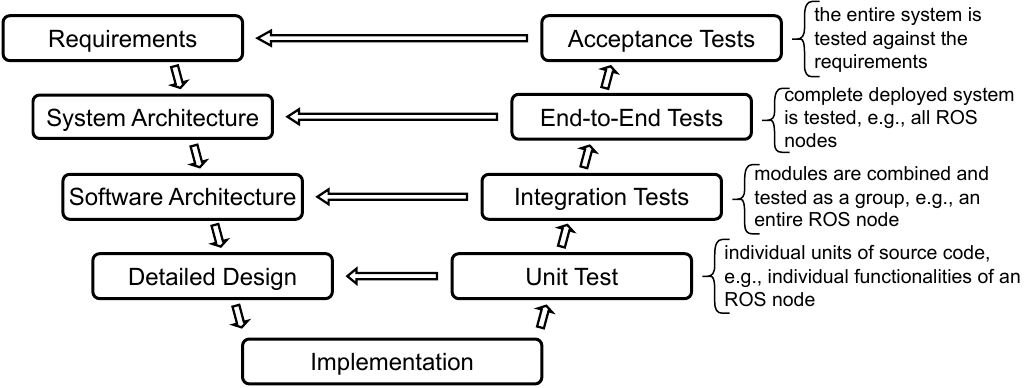}
	\vspace{-.3cm}
	\caption{Test levels in the V-model and examples of related parts of a robotic system}
	\Description{A diagram of a V-model with examples of the applicability to robotic system on the right-side. The test levels and examples are discussed in the full text.}%
	\label{fig:vmodel}
	\vspace{-.6cm}
\end{figure}

\looseness=-1
Test levels are best illustrated with the V-model. Depicted in \cref{fig:vmodel}, it is a software-development process model that emphasizes testing and that is widely used in safety-critical domains, including robotics\,\cite{lim2010ijsea, biggs2013experiences}. It highlights the connection between development activities (left-hand side of the V) and testing activities (right-hand side of the V).
Each development activity has a direct testing counterpart that evaluates the system against the development specification\,\cite{ruparelia2010software}. The V-model is only one process model; real development processes in practice combine different models, especially with iterative and incremental process models. Many agile processes are also iterative and incremental, and emphasize testing, especially automated testing.
These processes are often guided by model-driven and component-based design, which focus on reuse, formalization, and validation throughout the system lifecycle\,\cite{schlegel2009robotic, schlegel2010design}. Here, higher-level descriptions, i.e., models of how the system works (its structure and behavior), are created first. These models then drive or automate implementation, helping manage complexity and easing updates since code can be regenerated instead of rewritten.

\looseness=-1
\citet{lim2010ijsea} proposed a hierarchical test model, i.e., an organized set of testing levels (unit, integration, and system) for robot software components integrated with hardware modules. Their model is based on the V-model and uses black-box testing to show efficiency on an embedded robotic testbed. 
\citet{nagrath2021smartts} presented SmartTS, a component-based and model-driven method that automates testing at all levels and produces verifiable test records (i.e., results obtained by checking component behavior against model-derived contracts) to increase trustworthiness.

\parhead{Unit Tests}
\looseness=-1
At the lowest level of the V-model, unit testing verifies the correct implementation of small units, such as code functions. It focuses on each unit individually to ensure proper behavior. Unit testing corresponds to the activity \emph{detailed design} of the V-model, where each unit is specified and then verified against the specification.
Unit tests can be conducted early in the development. They can, in principle, be created for each code function, and testing early allows identifying and fixing faults early, which lowers engineering costs\,\cite{mcconnell1998software}.

\looseness=-1
In ROS-based systems, unit tests examine small fragments of it, such as individual functions, rather than testing a whole node itself\,\cite{bihlmaier2014simpar}. This level of granularity is consistent with general software testing practice, where unit tests target isolated functionality. To achieve this, the ROS documentation recommends decoupling core business logic from ROS-specific code (e.g., callbacks or communication handles) to facilitate testing with standard frameworks like \code{gtest} for C++ or \code{unittest} for Python. Furthermore, it advises writing tests to reproduce bugs prior to implementing fixes and replacing hardware with simulators or mocks to ensure deterministic and repeatable results\,\cite{ros2testing,rosunittest}.

\looseness=-1
To facilitate automated testing in robotics, \citet{bihlmaier2014simpar} proposed the Robot Unit Testing (RUT) methodology. 
RUT uses sensor and actuator data-based simulations within the ROS framework to test software components and robot behaviors.
This ensures high automation, independence, and parallel execution.
\citet{paikan2015generic} presented a generic framework for developing and running unit tests independent of programming language and middleware.
Tests are implemented as plugins, allowing easy reuse and grouping into suites. 
A fixture manager sets up the required environment, such as simulators or robot drivers, and monitors resources to ensure consistent test execution.

\parhead{Integration Tests}
\looseness=-1
These determine whether individual units of the system interact correctly with one another, allowing developers to build confidence in their combination, while verifying that their implementations meet requirements. While unit tests target methods or functions, integration tests focus on more coarse-grained components, such as an entire ROS node. 

\looseness=-1
In ROS, integration testing corresponds to the architecture level of the V-model. There, tests verify that modules (e.g., methods or functions) work together to form a functioning node and that multiple nodes interact properly to provide the actual functionality (e.g., topics, services, actions). For example, a perception node may publish data that is consumed by a planning node. According to the documentation, developers can use \code{rostest} (ROS~1) or \code{launch\_testing} (ROS~2) to run nodes in controlled scenarios. Common issues uncovered at this stage include mismatched message types, incorrect service interfaces, race conditions, and resource conflicts\,\cite{ros2testing,rosunittest}.

\citet{brito2020springer} proposed a graph-based method that represents publish/subscribe interactions as communication graphs, combining functional (black-box) and structural (white-box) testing to detect message-level faults, such as loss, duplication, and corruption. Their experiments showed significant gains in defect detection and coverage. Additionally, \citet{piel2010automating} presented two methods of automated integration testing of publish/subscribe networks. One generates a large number of event sequences to find a malfunctioning system state, and the other uses predefined event sequences to verify the behavior. \citet{kang2015automatic} addressed the test oracle problem for composite components. They developed an automatic generation algorithm, which uses I/O relationship analysis and atomic component test cases. Their method significantly reduced the time-consuming and error-prone task of manually generating expected results.





\parhead{End-To-End Tests}
\looseness=-1
These verify that the system has been built correctly by evaluating the whole robot system\,\cite{araujo2022tse}. At this level, all nodes are executed together and tested under conditions resembling real operation\,\cite{breitenhuber2020towards}. End-to-end tests are the highest verification level before deployment; they correspond to checking the system architecture at the left of the V-model. In ROS systems, these tests consider the complete software and all hardware parts, which are assessed in controlled scenarios to confirm expected behavior. They are often executed automatically, so developers can repeatedly exercise realistic behavior and detect issues that only emerge when all nodes interact\,\cite{erich2019design}.

Several works have explored end-to-end testing in robotics. \citet{erich2019design} presented an open-source framework that allows automated testing of collaborative robot applications and evaluates the behavior of the system while integrating with CI pipelines. \citet{breitenhuber2020towards} presented a behavior-driven framework for ROS that supports test-first development, introduces behavioral specifications, and enables automated end-to-end tests. \citet{cardoso2020towards} proposed a method for verifying modular robotic systems and guiding tests of individual components to build confidence in the complete system. \citet{cossellnovel} showed how validating software environments and configurations in multi-agent platforms can improve system setup and simplify troubleshooting.

\parhead{Acceptance Tests} These tests validate that the right system has been built by checking whether the robotic system meets its predefined requirements\,\cite{araujo2022tse, nguyen2025}. 
The whole system is evaluated against a specific operational scenario composed of several features, confirming that the robot delivers the behavior expected by users or stakeholders.
As in the previous level, the system is assessed in its entirety, including input and output from external sources such as obstacles or sensor noise. 
Acceptance tests reflect the validation activities on the left side of the V-model, where the system is checked against the requirements\,\cite{santos2022irc, santos2024frontiers}. This stage requires confirmation by users or customers and provides assurance that the system is suitable for its intended purpose.

\looseness=-1
Several studies have explored acceptance testing in robotics. \citet{estivill2018continuous} tested full robotic behaviors in simulation using a simplified Model–View–Controller (MVC) pattern for GUI-stripped (graphical user interface) simulations, which allowed for efficient assessment of complex missions, sometimes even faster than in real life. \citet{kramer2006ugv} proposed an automated procedure for customer acceptance of unmanned ground vehicles, which exercised all capabilities over extended periods in standardized environments and uncovered numerous failures. \citet{nguyen2025} advanced behavior-driven development further by adding timing and system behavior modeling, enabling the creation and automatic simulation-based checking of specific acceptance criteria. Santos et al.\,\cite{santos2022irc,santos2024frontiers} proposed converting business requirements into executable scenarios, demonstrating reliable fault detection and improved software quality for industrial robotic systems.

\begin{tcolorbox}[myexample]\looseness=-1
	For the delivery task (cf.~\cref{tab:requirements}, requirement~10), unit tests could focus on single nodes, such as the velocity controller or the obstacle detection component, using simulated sensor inputs to verify behavior in isolation. Integration tests would then combine nodes, for instance, navigation, obstacle avoidance, and trajectory tracking, to ensure they interact correctly when the robot plans and executes a path. Next, end-to-end tests could execute the full system, observing the robot moving from the kitchen to a customer table, while avoiding obstacles and following safety constraints. Finally, acceptance tests would place the robot in realistic restaurant conditions with dynamic obstacles and human interactions to test the whole system. 
\end{tcolorbox}

\subsection{Test Types}\label{sec:sub:test-types}
\looseness=-1
Test types define what is tested and give meaning to testing\,\cite{iso29119pt1}, whether targeting functional correctness, performance, or other quality attributes, such as safety and reliability. 
That is, they are divided into functional and non-functional testing. 
Functional testing targets the functionalities of a system, that is, verifying an operational or general property of a system. 
Non-functional testing targets the quality of the system under different operational conditions\,\cite{hass2014guide, istqb2018, baresi2006introduction}, and is divided into two categories: quality, which addresses the quality of the system, and constraints, which address, e.g., external regulation, or safety standards the system has to comply with\,\cite{glinz2007non}.   


\subsubsection{Functional Testing}
\looseness=-1
Functional testing evaluates the completeness and correctness of a system's specific functionalities. Therefore, functional tests are derived from the system's functional requirements and are designed to verify that each component complies with them.
In robotics, functional testing often focuses on confirming that the robot can correctly perform the tasks or actions it was designed for. That is, focusing on evaluating obstacle avoidance and navigation performance in dynamic environments\,\cite{paez2022pedestrian}, path-following accuracy and motion stability in mobile robots\,\cite{chowdhury2018experiments, kamandar2022design, allozi2022feasibility}, and perception systems for visual feedback control in industrial robots\,\cite{korayem2005vision}. Furthermore, functional testing assesses the end-effector capabilities, including grasp strength, repeatability, and cycle time\,\cite{falco2020benchmarking, liu2021ocrtoc, kimble2020benchmarking} for manipulator systems. In networked robotic applications, functional tests include verifying communication reliability and latency in distributed control scenarios\,\cite{satoh2019developing}.

\looseness=-1
With the rapid growth of robotic systems, efforts to define functional tests are conducted, aiming to make tests comparable and standardized across application domains\,\cite{jacoff2010comprehensive, norton2019standard, hietanen2021benchmarking}. That includes establishing agility benchmarks for multi-legged robots\,\cite{eckert2019benchmarking}, navigation and obstacle evaluation standards for UAVs\,\cite{yoon2019analysis}, and task-based performance measures linking pose estimation errors to task success in manipulation\,\cite{hietanen2021benchmarking}. 
In summary, functional testing in robotics means determining functional correctness, combining perception, control, and environment interaction.

\begin{tcolorbox}[myexample]\looseness=-1
Functional testing could mean evaluating the navigation and obstacle avoidance capabilities, e.g., verifying requirements 10 from \cref{tab:requirements}. The goal would be testing whether the robot can reach a destination without collision, maintaining localization accuracy and adapting to environmental changes between its internal map and the physical layout.
	Verifying functional correctness could rely on the standardization procedure by\,\citet{norton2019standard}. The robot would repeatedly navigate between two locations A and B in a bounded environment containing configurable static and dynamic obstacles.
	Performance metrics (e.g., task success rate, average traversal time, collisions) would be recorded by external observation. Applied to our restaurant robot, we define two locations, e.g., Kitchen (A) and Table 3 (B), and add obstacles, e.g., chairs or moving humans, to simulate reality. The robot would navigate from A to B, updating its map upon unexpected obstacles. Functional correctness is determined by measuring task completion time, success rate over repetitions, and trajectory deviation.
\end{tcolorbox}

\subsubsection{Non-Functional Testing}
\looseness=-1
Non-functional testing focuses on aspects that are not directly related to the functionality, but rather the conditions and qualities under which the system operates, such as performance constraints, and usability expectations\,\cite{iso29119pt1}. 
The term is not precisely defined in the literature and often encompasses under-specified functional requirements.
To address this, non-functional requirements can be further divided into qualities and constraints\,\cite{glinz2007non}.
The former relates to measurable properties of system performance under various conditions, such as usability, or resource utilization. The latter relates to compliance with external rules or regulations set by stakeholders, regulatory bodies, or industry standards.




\parhead{Quality Testing}
Evaluating how well a system performs under various operational conditions, rather than correct behavior, is called quality testing. It assesses quality characteristics such as performance under load, scalability, and compatibility, against defined quality requirements\,\cite{bath2014}. The goal is to determine how the system behaves with changing workloads or environmental factors. 
For example, a test can focus on changes in response times with increasing computational load, and whether performance remains within acceptable limits. 
Quality tests provide evidence that the software can operate reliably and efficiently across a range of expected conditions. 

\looseness=-1
\textit{Usability Testing} evaluates the user experience with a system, including intuitiveness and efficiency. 
Usability is particularly relevant for human-robot interaction, where the quality of interfaces influences user acceptance and task performance. 
\citet{golchinfar2023let} studied the usability of a GUI for remote-controlling a service robot during exceptional situations and reports lessons learned for designing an interface that improves remote human intervention.
\citet{casiddu2020humanoid} proposed guidelines for designing usability tests for robots, addressing aspects such as participant recruitment, sample size, session duration, invasiveness, and influence of the test environment. \citet{moon2017usability} developed quantitative usability evaluation criteria for movement-support robots designed for the elderly, combining interviews and empirical assessment to measure the interaction quality.




\textit{Performance Testing} evaluates how efficiently and reliably a system performs its intended functions when operating under typical or expected load\,\cite{iso29119pt4}. 
Tests can consider varying operational conditions with relation to system performance, such as responsiveness, resource utilization, and capacity\,\cite{iso29119pt1}.
\citet{wienke2017ar} introduce a framework to specify, execute, and analyze performance tests for individual components of robotic systems. The component under test is integrated into a controlled test environment that generates input data and load conditions while monitoring performance metrics. 
That is, tests measure how quickly a system responds to user requests, how resource consumption scales with concurrent operations, or whether performance degrades under peak demand conditions. 
It should be made clear that accuracy is a functional requirement, whereas, responsiveness and load handling are non-functional performance-related requirements. 
At the system level, \citet{okwu2022development} conduct performance testing of an autonomous lawn mower by measuring the time to task completion for different grass heights, illustrating how environmental factors influence robotic performance. Similarly, \citet{gertman2007methodology} propose a method for evaluating the performance of human-robot teams. 
Performance tests are often subdivided into more specific test types, such as load, stress, endurance, and scalability testing, depending on the intensity of workload\,\cite{iso29119pt4}, that we consider as separate test types.




\textit{Load testing} evaluates performance under varying loads, typically ranging from low and nominal to peak loads\,\cite{iso29119pt4}.
The goal is to determine the break point, that is, at which point the system performance becomes unstable and unresponsive under realistic workload conditions. 
For example, determining how many simultaneous users, processes, or sensor inputs can be handled before failures occur.
We did not find papers that explicitly target load testing in the robotics domain.

\textit{Stress testing} evaluates the performance when it is pushed beyond its anticipated peak load or when available resources (e.g., memory, processor, disk) are reduced below specified minimum requirements, to evaluate how it behaves under extreme conditions\,\cite{iso29119pt4, bath2014}. Stress testing pushes the system to its absolute limits and evaluates the behavior beyond the break point. 
\citet{collet2019stress} introduces RobTest, a tool for stress testing single-arm robots by maximizing the load of the CPU.
Stress testing for software systems also encompasses volume testing\,\cite{iso29119pt4}, which evaluates the performance when processing large volumes of data or when data storage approaches capacity limits. However, we did not find papers that target volume testing in the robotics domain.


\looseness=-1
\textit{Endurance testing} evaluates the ability to maintain performance and reliability over an extended period of continuous operation\,\cite{iso29119pt4}. The objective is to detect issues such as resource leaks or gradual performance degradation that may not be encountered in shorter tests. In robotics, endurance testing is typically performed by operating a specific subsystem, such as perception, navigation, or communication, or the entire system continuously over extended time periods\,\cite{broggi2013extensive,arango2020drive}. 
\citet{ball2016vision} conducted extensive field tests over several weeks to evaluate the robustness of an obstacle detection and navigation system under real conditions. \citet{babic2020vehicle} developed a vehicle-in-the-loop test environment to enable long-term testing of autonomous driving systems, allowing continuous evaluation of autonomy performance without the safety and cost constraints of physical field deployment. 






\looseness=-1
\textit{Capacity testing} (also called scalability testing) evaluates performance under conditions that may need to be supported in the future. This includes evaluating what level of additional resources (e.g., memory, disk capacity, network bandwidth) will be required to support anticipated future loads\,\cite{iso29119pt4}. It ensures that the system can handle increasing loads, such as higher data volume, user load, or computational demands, without compromising future performance. 
Capacity testing in robotics often addresses if the software scales with an increasing number of robots. \citet{humphrey2007assessing} developed a scalable interface for a human-multiple robot system, focusing on the effects of increasing the number of robots on workload, situation awareness, and robot usage. \citet{hamann2021scalability} proposed a general model of scalability for multi-unit systems and applied it to multi-robot systems.

\looseness=-1
\textit{Resource utilization testing} evaluates the efficiency of computational and energy resources usage under various operating conditions. It typically measures CPU and GPU load, memory usage, and network bandwidth to identify bottlenecks, ensuring that resources are allocated effectively at runtime. 
These measurements help verify that software systems meet performance and efficiency requirements without excessive resource consumption. In robotics, however, resource utilization testing ensures real-time constraints are met, even during computationally complex tasks.

\citet{maruyama2016emsoft} discuss performance and resource utilization aspects in ROS2, highlighting the challenges of ensuring predictable timing behavior. Similarly, \citet{bedard2022iral} integrated tools such as LTTng into ROS environments to monitor CPU and memory usage during testing. \citet{wienke2018irc} proposed a DSL (cf.~model-based testing in~\cref{model-based-testing}) to systematically specify performance and resource utilization tests for individual robotic components, enabling parametrized test cases that capture CPU, memory, and network demands over different execution phases.
Compared to traditional software, energy consumption is a critical resource in robot software. \citet{swanborn2021icse} developed Robot Runner, a tool to automate experiments to evaluate energy consumption and CPU/memory use. \citet{touzout2022mobile} extended it by integrating an energy consumption model of a differential-drive robot into a ROS/Gazebo simulation environment, improving monitoring and energy management. \citet{gadaleta2021extensive} analyzed energy consumption of high-payload industrial robots, measuring energy during tasks at different load conditions to validate virtual prototypes. 
A special case of resource utilization testing mentioned by software standards is memory management testing, focusing on performance evaluation in terms of memory usage\,\cite{iso29119pt4}. However, we did not find robotics papers reporting on memory management testing.




\looseness=-1
\textit{Reliability Testing} evaluates whether the system can consistently perform its intended functions over time with varying conditions. The goal is to uncover failures or unexpected behaviors that may not appear during short-term operation, focusing on stability, dependability, and frequency of failures. Unlike endurance testing, which emphasizes sustained operation and general performance over time, reliability testing measures metrics such as mean time to failure or mean time between failures, assessing the likelihood that the system will function correctly for a specified period under defined conditions\,\cite{iso29119pt4}.
In robotics, reliability testing typically involves long-term simulations, repeated execution of scenarios, and real-world experiments, often complemented by systematic fault injection to assess the system's robustness under abnormal conditions\,\cite{hereau2021fault, wang2020reliability}. \citet{tang2011testbed} developed a dedicated testbed for evaluating real-time prognostics, health management, and automated contingency management techniques in autonomous vehicles, enabling systematic reliability assessment. \citet{Deng01012021} proposed a timed-automaton-based method to record the occurrence of reliability-related events in automated guided vehicles, providing a quantitative measure of system dependability. Similarly, \citet{hereau2021fault} introduced a fault-tolerant control architecture for underwater robots, where undesirable events are detected and managed by a fault-tolerant module that either recovers the system or reports to a mission manager, thereby improving overall system reliability.

\looseness=-1
\textit{Security Testing} evaluates whether a system contains vulnerabilities or flaws that could compromise integrity or confidentiality. The goal is to identify vulnerabilities that could allow unauthorized access, tampering, or manipulation of the system or its data\,\cite{iso29119pt4}. In robotics, security testing is particularly critical due to the interaction of robots with physical environments, networks, and human operators.
\citet{fernandez2019securing} employed the Elliptic Curve Integrated Encryption Scheme to protect data transmissions from malicious attacks for UAV communication systems. \citet{reithner2021analysis} conducted penetration tests on mobile robots
, revealing vulnerabilities that could compromise both robotic operation and network safety. Security testing also targets middlewares such as ROS, for instance, \citet{sim2023sminer} developed SMINER to identify vulnerabilities arising from unrestricted or falsely implemented behaviors within ROS.
Beyond tools and frameworks, security testing in robotics benefits from structured engineering practices. 
\citet{mayoral2022iros} introduced SROS2, a set of usable cybersecurity tools and libraries for ROS 2 that help developers add authentication, authorization, and encryption capabilities to robotic systems. Simplifying the process of introspecting ROS 2 computational graphs, generating and deploying Identity and Permissions Certificate Authorities, and converting ROS 2 security configurations into DDS permissions. They proposed a six-step DevSecOps-based methodology for securing ROS2 computational graphs, encompassing modeling, artifact generation, and continuous monitoring.

\looseness=-1
\parhead{Constraints Testing}
Tests evaluate a system's adherence to external regulations or requirements set by stakeholder, regulatory bodies, or industry standards\,\cite{iso29119pt4}. Constraints can address various topics, including safety regulations, data privacy and protection, insurance policies, and laws.

\looseness=-1
\textit{Compliance Testing}
aims to verify that the system adheres to specific standards, regulations, or legal requirements and is also known as conformance or regulatory testing.
In robotics, the absence of widely adopted, domain-specific standards typically means systems must comply to the regulations applicable to their operating environment. 
For example, autonomous vehicles must conform to traffic laws and safety requirements\,\cite{bi2020rcim}. \citet{bensalem2005testing} proposed checking compliance against formal specifications on traces generated by an instrumented system for the automated testing of real-time application conformance.
\citet{sohail2023case} integrated simulation-based and property-based testing to verify robot standard conformance, formalizing and automating the execution of requirements defined in robotics standards as reusable simulation scenarios.
This enables partial automation of conformance testing across diverse robot platforms and reveals undetected software and driver defects.




\textit{Safety Testing} focuses on ensuring that the system does not, under defined conditions, cause injury, harm to humans, or damage to property. Safety testing can be considered a subset of compliance testing, but unlike compliance testing, it often lacks universally hard-defined rules, making its evaluation context-dependent. In robotics, safety is particularly critical when humans share the operational environment with robots or interact closely in collaborative tasks\,\cite{gleirscher2022verified, huck2021testing}.
\citet{gleirscher2022verified} developed a safety controller for a collaborative manufacturing, combining synthesis, formal verification, and testing to ensure safe human-robot interaction. \citet{huck2021testing} formalized hazard analysis as a search problem and proposed simulation-based strategies to generate test cases that identify hazardous behaviors due to unexpected human actions. For legged robots, \citet{weng2022safety} propose sampling of scenarios to characterize overall safety performance, systematically testing potential failure modes under a variety of operational scenarios.





\begin{tcolorbox}[myexample]\looseness=-1
To evaluate our restaurant robot's resource (e.g., CPU) utilization we could use the framework of \citet{wienke2018irc}.
Using their DSL, developers could model repeated path planning requests or simulated camera streams, while the framework monitors CPU and memory usage under different input rates or environment complexities. For instance, the navigation stack should maintain CPU utilization below a defined threshold to prevent delays in motion control or task management. The framework generates synthetic test inputs simulating dynamic obstacle data from laser sensors and measure how the navigation algorithm processes them.
\end{tcolorbox}

\subsection{Test Activities}\label{sec:sub:test-activities}
\looseness=-1
Having established the meaning of test environment, test levels, and test types, the next step is to structure them in a meaningful way, that provides structure to the complete testing of what to test, how to test it, and when to test it\,\cite{norris2019system}.
Test activities are steps carried out to perform testing throughout the test plan. These steps exist regardless of test plan, or strategy they define how testing is executed.
We will not discuss test process, test plan, and test strategy in detail, as they are very individual for each project or organization, but we provide an intuitive understanding.
The test process and test plan provide the framework and guidance for organizing the test activities. A test plan is a detailed document that defines the test objectives (what should be achieved with testing), the strategies and resources for achieving them, and the schedule for executing tests, thereby coordinating all testing efforts. The test process, in turn, consists of a set of systematic test activities carried out to fulfill the test objectives\,\cite{iso29119pt4}. The test strategy is a high-level, organizational or project-level description of how testing will be approached. It defines the principles, general methods, and rules guiding all testing work.
Research in this direction considers optimization methods, aiming to minimize total testing time while ensuring full coverage of test items, often under resource constraints\,\cite{tang2020test, gotlieb2020testing, looije2017specifying}. 
 

\parhead{Requirements Analysis and Test Planning} 
This activity involves analyzing the system requirements, deriving test objectives, and selecting appropriate test types. By the end of this phase, a detailed plan is produced that specifies which parts of the system are tested, how they are tested, and against which requirements the testing is performed.



\looseness=-1
\textit{Analyze Requirements} identifies all functional and non-functional requirements relevant to testing. These requirements are examined in detail, together with other relevant artifacts, such as design specifications, behavioral models, and source code. Together they form the test basis, from which the test objectives and test items---the specific units, components, or subsystems to be tested---are derived.
This step ensures that the subsequent testing activities are systematically aligned with the intended system behavior and quality characteristics.


\looseness=-1
\textit{Test Planning} defines test conditions (i.e., what should be tested) and test types---5
for each test objective and item from the requirements analysis
\,\cite{iso29119pt1,istqb2018}.
Since it is often impractical to test entire systems exhaustively, test planning includes prioritizing which components require more thorough evaluation. Finally, this phase determines the most suitable test types (see \cref{sec:sub:test-types}) to achieve the defined test objectives, establishing a structured roadmap for subsequent test design and execution.







\parhead{Test Design and Implementation}
Appropriate test design techniques are applied to systematically derive or select test cases based on the defined test condition and test types.
During this activity, test conditions are transformed into concrete, executable test cases.
When designing the test suite, i.e., the complete collection of test cases, coverage objectives are established to ensure adequate verification of the system under test.
Test coverage is defined as the degree to which coverage items derived from the test basis and test conditions are exercised by the test suite.
Coverage may include specification-based coverage (e.g., requirements coverage), structural coverage (e.g., statement, branch, or condition coverage), as well as coverage of test data variations and execution environments. Traceability between test cases and test basis is maintained to support coverage analysis and impact assessment.
It should be noted that coverage criteria serve as indicators of test completeness, but do not in themselves guarantee defect detection effectiveness.
This phase results in a detailed test suite, together with defined coverage criteria and traceability information, providing the foundation for systematic test execution and evaluation.





\looseness=-1
\parhead{Test Environment Setup and Execution}
Once the test suite has been designed, it is executed within a controlled test environment that reflects the preconditions defined for each test case. The test environment may consist of a simulation-based setup, or a real-world setup. The selection depends on various factors, such as the test objectives, and the risks associated with execution. 

\looseness=-1
During execution, relevant data such as performance metrics, and system states are recorded for analysis. In robotics, continuous integration and continuous deployment pipelines are uncommon. However, means to improve are still conducted, for example, \citet{swanborn2021icse} introduced Robot Runner, a tool designed to streamline the execution of measurement-based experiments by automating experiment setup, execution, and replication for both simulated and physical robots. 



\looseness=-1
\parhead{Test Evaluation}
Following the test execution phase, the collected data is analyzed and compared against the expected outcomes to determine whether each test has passed or failed. This evaluation assesses whether the observed behavior aligns with the specified requirements and whether the defined test objectives have been met.
However, not only individual test results are evaluated, but also a comprehensive evaluation of test coverage is conducted, ensuring that all requirements were covered by the test suite.
The evaluation phase provides feedback on whether the system meets the requirements, deviations or failures may indicate design flaws, implementation errors, or inadequate testing.
If issues are detected, either the code or the test suite needs to be adapted. Any changes to the code base result in regression testing, ensuring that the changes do not introduce new faults.

\begin{tcolorbox}[myexample]\looseness=-1
	We illustrate the activities using requirement 10 (cf.~\cref{tab:requirements}) as part of the test basis.
	First, the \textbf{analysis} of the requirement determines the purpose of testing and which components are involved.
	For requirement~10, a test objective would be to verify that the robot navigates to target locations while avoiding collisions and complying with safe velocity limits.
	The corresponding test items would be the components Motion planning \& control and SLAM in the component control layer.
	From the test objective, test conditions representing testable aspects of requirement 10 could be derived, which include, (i) computing a collision-free path to the target, (ii) detecting and avoiding dynamic obstacles, (iii) enforcing a maximum velocity of 0.8\,m/s, and (iv) stopping the robot within the tolerance distance to the target.
	Second the \textbf{planning} based on the test conditions and objective, leads to appropriate test types.
	For requirement 10 this includes, functional testing to validate correct task execution, performance testing to measure traversal time and velocity, and safety testing to ensure compliance with speed and collision constraints.
	Thereafter, specific test cases would be \textbf{designed and implemented}.
	A test case would be:
\begin{leftbar}
	\textit{Precondition}: The robot is powered on, localized in the kitchen area, and the restaurant map is loaded into the navigation stack.
	
	\textit{Test Input}: A navigation command to move to Table~3, provided in the form of the table's coordinates in the restaurant.
\end{leftbar}
\begin{leftbar}
	\textit{Expected outcome}: The robot plans a collision-free path to Table~3, navigates while avoiding dynamic obstacles that appear during motion, and never exceeds 0.8\,m/s.
	
	\textit{Postcondition}: The robot reaches Table~3, stops within 0.3\,m of the table, and signals task completion to the mission manager.
\end{leftbar}
\looseness=-1
	During \textbf{test environment setup and execution}, the tests are executed, either in a real restaurant or in simulation. For our example, a Gazebo simulation environment of the restaurant, with logging and monitoring tools, capturing the robot's position, velocity, and task states needs to be developed.
	Finally, in \textbf{test evaluation}, the test results (e.g., trajectory/velocity logs, and delivery completion timestamps) would be compared against expected results. Deviations, such as collisions, overspeeding, or missed delivery confirmations, would be recorded as failures and followed by regression testing to verify corrective changes.
\end{tcolorbox}

\section{Test Design Techniques}\label{sec:sub:test-design-techniques}

We have discussed how to structure the test process and give meaning to each test. We now discuss techniques to systematically derive test cases from different sources of information.
These test design techniques classify into: \emph{Black-box testing}, also known as specification-based techniques, \emph{White-box testing}, also known as structured-based testing, and \emph{Experience-based testing}\,\cite{iso29119pt4, khan2012comparative}.

\begin{figure}[b]
\vspace{-.6cm}
  \includegraphics[clip, trim=0.5cm 2.9cm 0.5cm 8.0cm, width=\textwidth]{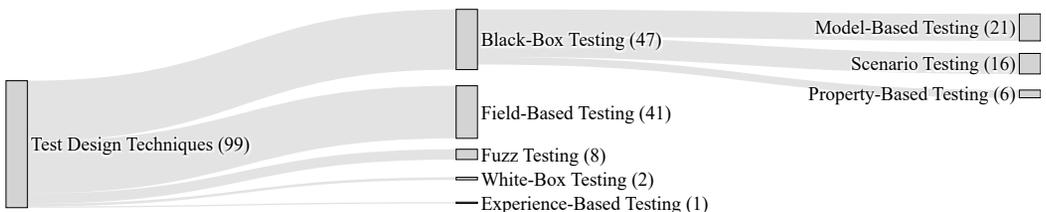}
	\vspace{-.8cm}
  \caption{Number of robotics papers reporting about the respective testing design techniques}
  \Description{A Sankey diagram that visualizes the classification of papers discussed in this survey.}
  \label{fig:subsectionPapers}
	\vspace{-.5cm}
\end{figure}

\looseness=-1
We found that the majority of papers focus on black-box testing (47), while white-box testing is nearly absent (2). Field-based testing has the second most papers (41), followed by fuzz testing (8), and finally, experience-based testing (1). Among the black-box testing techniques, model-based testing (21) and scenario-based testing (16) are most prominent. \Cref{fig:subsectionPapers} shows the distribution.

\subsection{Black-Box Testing}
In black-box testing the system is treated as a black box, that is, test cases are derived from externally observable or specified behavior of the system, instead of internal structures.
The behavior can be captured through models, requirements, or other forms from which test cases can be derived.
In robotics, black-box testing is widely used, whether to test the robustness of the system by generating test cases with mutating valid sequences of requests\,\cite{powell2012testing}, or a web-based black-box testing framework to test components of robotic software\,\cite{kang2012web}. 



\looseness=-1
\textit{Equivalence Partitioning}
is a technique that divides input and output data into equivalent classes or partitions. A partition is a set of input or output data that is treated the same by the system. Testing one input from a partition is assumed to be representative for the whole partition\,\cite{istqb2018,iso29119pt4}.
\citet{weinstein2024equivalent} introduce equivalence classes of environments to capture situations in which a robot cannot distinguish between different worlds based on its control signals and sensor outputs. These equivalence relations group environments into classes that appear identical to the robot.


\looseness=-1
\textit{Boundary Value Analysis} focuses on identifying errors at edges or boundaries between partitions (often derived using equivalence partitioning). 
For each boundary, testing focuses on values immediately on either side of the edge---both within and just outside the equivalence class\,\cite{istqb2018,iso29119pt4}. 
We did not find any papers reporting on boundary value analysis in robotics.





\looseness=-1
\textit{Classification Tree Methods} group test inputs, like equivalence partitioning, into non-overlapping classes using a tree representation. 
Each classification represents relevant aspects to test, e.g., input types or system states\,\cite{iso29119pt4}. We did not find any papers reporting this method in robotics.

\looseness=-1
\textit{Combinatorial Testing} obtains a meaningful and manageable subset of test cases, often employed to test the interactions of numerous parameters with discrete values. Selecting a combination of parameters with specific values reduces the number of test cases. A combination is defined as parameter $P$ and values $V$ the parameter can take.  
Test coverage is derived by selecting a subset of $P$-$V$ pairs. Different techniques for deriving combinations of $P$-$V$ exist, for example, all unique combinations, pair-wise combinations, the set of $P$-$V$ pairs that have value of $V$ at least once (also known as each choice testing), or the set of $P$-$V$ pairs resulting from fixing all parameters to a ``base'' value and then varying the value of one parameter at a time, setting it to its non-base valid values (known as base choice testing)\,\cite{iso29119pt4}. 
\citet{kluck2023empirical} used combinatorial test design techniques to create and manage scenarios for advanced driver assistant systems by using ontologies to describe the environment.


\looseness=-1
\textit{Decision Table Testing} evaluates combinations of conditions to determine distinct outcomes, particularly effective for logical relationships. These tables are constructed by defining system conditions and corresponding actions. Each column represents a decision rule, a unique condition combination, and its associated actions. In limited-entry decision tables, condition and action values are typically Boolean. Conversely, extended-entry decision tables may take multiple values for conditions and actions, such as numerical ranges, equivalence partitions, or discrete values\,\cite{istqb2018,iso29119pt4}. We did not find any papers reporting on decision table testing in robotics.


\parhead{Model-Based Testing (MBT)}\label{model-based-testing}
\looseness=-1
MBT systematically derives test cases from behavioral models that define the required or expected behavior of a system and should not be confused with environment or simulation models,\,\cite{iso29119pt4, istqb2018}.
These models abstract different aspects of the system, targeting the full system or individual components\,\cite{ernits2015emcr, proetzsch2010lcns}, or capture the behavior in different contexts, such as human-robot interaction\,\cite{micskei2012amsta}, and support systematic test case generation by exploring the model's states and transitions\,\cite{kanter2020rie, micskei2012amsta}.

Common representations can be formal or semi-formal, include state machines, automata, or DSLs\,\cite{dslbook} which capture the expected behavior and interactions\,\cite{honfi2017sdl}.
%
State transition testing, a special case of MBT, models the system as a state machine, capturing all relevant states and transitions\,\cite{iso29119pt4,istqb2018}. Inputs cause changes in system states, and test cases are designed to cover valid and invalid transitions between states. A test case is usually represented as a sequence of inputs/events, which results in a sequence of state changes (and actions, if needed).

\looseness=-1
MBT has been widely adopted in robotics, particularly because behavioral models leverage complex interaction descriptions, focusing on different aspects, such as safety, robustness analysis, interaction (human-robot or multi-agent), corner-case exploration, and coverage.
%
\citet{micskei2012amsta} presented context models that model the behavior for complex situations for safety and robustness of the control software of autonomous systems, allowing context-aware scenario generation, from which test cases can be derived.
\citet{zendel2013vitro} introduced VITRO a framework to evaluate the robustness of computer vision solutions by automatically creating 3D scenes from domain descriptions and creating test cases from them. 

\looseness=-1
%
For collaborative tasks, models are used to define interaction behavior and generate test cases, such as in human-robot interaction\,\cite{webster2020corroborative}, or autonomous driving\,\cite{sarkar2019behavior}. Colored petri nets have been used to model interactions and system dynamics, supporting verification of the cooperative behavior\,\cite{lill2014testing, saglietti2016model}. Test-case generation is guided through coverage criteria of the petri nets and evolutionary algorithms to maximize interactions while minimizing the number of test cases\,\cite{saglietti2016model}.
In human-robot interaction, Belief-Desire-Intention agent models emulate intelligent and adaptive robots, providing rational agency and a reasoning mechanism for planning, to support realistic test generation\,\cite{araiza2016intelligent}.

\looseness=-1
DSLs\,\cite{dslbook}, or other tools and frameworks, support the systematic generation of test cases, including corner case identification. 
For instance, graphical DSLs have been used to specify models as graphs of locations and transitions, annotated with probabilities and events, enabling the automatic generation and execution of test cases in simulation environments\,\cite{proetzsch2010lcns}.
The TestIt framework supports integration testing of ROS packages, by providing an interface for model checking tools like Uppaal\,\cite{kanter2020rie, kanter2019dessert}. It allows model generation from behavior trees or other specifications and includes the creation of placeholder components for dependencies, enabling testing of complex software.

\citet{nunez2023implementation} present a formal testing theory for probabilistic, cyclic robotic control software by introducing ptockLTS models and defining a comprehensive set of implementation relations that enable sound, systematic MBT under varying observability assumptions.
For modeling environmental behaviors and environmental factors, extended finite state machines have been used. Test cases are generated using graph-coverage criteria over abstract world models and then transformed into executable test cases using input-space partitioning\,\cite{andrews2015active}.


Further research aims to develop tools and frameworks, or integrate continuous integration into MBT workflows to automate test generation and execution\,\cite{weiss2012multi, wei2017evolving, honfi2017sdl, katz20202icra, khosrowjerdi2018learning, martin2023verification, mossige2015testing}.

\begin{tcolorbox}[myexample]\looseness=-1
  For our restaurant robot, we could evaluate the perception system, especially requirements 3 and 8, with MBT and the VITRO framework\,\cite{zendel2013vitro}. 
  First, we would define a domain model representing the restaurant environment including relevant objects (e.g., tables, human hands, trays). 
  Each object would be assigned visual properties (e.g., shape, size, color, texture) 
	and referenced with 3D models.
  Changing visual conditions, such as shadows, reflections, and glare from overhead lights, called ``visual criticalities'' in VITRO, could be selected from the provided catalog.
  From this domain model, VITRO could automatically generate 3D restaurant scenes and render synthetic images that systematically vary object configurations and lighting to achieve measurable domain coverage. For example, it could generate scenes where hands are partially occluded by other objects, or where motion blur affects hand gestures. Each rendered image would be paired with automatically generated ground truth annotations, such as object masks and bounding boxes for ``plate'' and ``hand'' Using clustering and evenly distributed sampling, VITRO reduces redundant test cases. 
  These generated datasets would then be input to the perception components to evaluate their performance under varying conditions. The automatically produced ground truth would allow evaluation of functional metrics (e.g., precision and recall) and robustness metrics (e.g., sensitivity to lighting changes or occlusions). Practically, the robot's camera parameters (field of view, resolution, lens distortion) would be mirrored in the VITRO rendering setup to ensure realistic input conditions.
\end{tcolorbox}

\looseness=-1
\parhead{Scenario Testing}
This method evaluates system behavior through realistic (often end-to-end) interaction sequences instead of isolated input/output pairs\,\cite{iso29119pt4}. 
Scenarios capture realistic situations that the system may encounter\,\cite{munoz2014first,straub2013characterization,carlone2009comparative}.
A scenario describes expected behavior across a flow of events, capturing both nominal (``main'' scenarios) and unexpected or failure (``alternative'' scenarios) cases.
In robotics, scenario-based testing is widely adopted, because it allows specifying sequential interactions, environmental context, and dynamic conditions easily.
Domain-specific languages are commonly used to model environments and define diverse test scenarios, allowing to specify complex scenes and environmental conditions in a structured format\,\cite{queiroz2019geoscenario, micskei2012amsta,queiroz2024driver}.
Such DSLs enable systematic and repeatable test scenario generation for simulation-based testing, which is particularly useful in domains like robotics and autonomous driving, where system behavior depends on a series of interactions and context, rather than isolated test cases\,\cite{queiroz2024driver}.

Applications of scenario testing in robotics range from validating manipulation and navigation behavior to evaluating robustness under uncertainty.
For example, the STAMINA project evaluated a robotic kitting system using six typical scenarios to assess performance under realistic conditions\,\cite{krueger2019testing}.
Scenario testing has also been applied to test the robustness of prognostics-enabled decision-making algorithms under sensor noise and timing disturbances\,\cite{sweet2014demonstration}, to reconstruct crash scenarios for testing autonomous driving algorithms from dashcam footage\,\cite{bashetty2020deepcrashtest}, and to validate UAV landing strategies on fixed or mobile platforms through three controlled scenarios\,\cite{guo2020precision}.
Similar, \citet{fornasier2021vinseval} introduce VINSEval, a simulation-based framework for systematically evaluating visual-inertial navigation systems (VINS) using parametrized scenarios with controlled variations in sensor noise, illumination, and environmental conditions.

\looseness=-1
A variety of tools support scenario generation and execution.
AmbieGen provides search-based scenario generation using evolutionary algorithms to identify critical scenarios\,\cite{humeniuk2022search, humeniuk2023ambiegen}. 
To test scenarios involving interactions between humans and robots, model-driven frameworks that automatically generate formal models and properties from configuration files and run statistical model checking experiments are deployable in real or simulated environments\,\cite{lestingi2021deployment}.
Scenarios can be generated for autonomous vehicles using imitation learning surrogates to guide an adaptive search, iteratively training an imitator agent to simulate the real agent and generate increasingly accurate predictions of scenario performance\,\cite{mullins2018accelerated}.
\citet{ortega2024frontiers} introduce composable and executable scenarios for simulation-based testing of mobile robots, along with tooling that enables developers to specify, reuse, and execute them. They allow adding new semantics, such as dynamic elements (e.g., doors) or varying task specifications, to existing models without modification, allowing the creation of rich, but manageable scenarios.
\citet{kang2022behavior} propose BTScenario, a scenario specification and test case generation tool for autonomous driving systems. BTScenario supports a high-level description of the spatial layout of map objects, such as roads and junctions, and a behavior-tree-based specification of the temporal behaviors of actors, including cars and pedestrians.
Additionally, frameworks using scenario models expressed as extended UML sequence diagrams offer structured robustness and functional safety evaluations, including detailed trace comparison and coverage metrics\,\cite{micskei2012amsta}.

\begin{tcolorbox}[myexample]\looseness=-1
 We could evaluate our restaurant robot's navigation and obstacle avoidance (requirement 10, cf.~\cref{tab:requirements}) with scenario testing and the AmbieGen framework\,\cite{humeniuk2023ambiegen}.
  %
  %
  In AmbieGen, the restaurant environment would be parametrized by its size, the minimum and maximum obstacle size, and position.
  It encodes these environmental parameters as individuals in a search-based optimization process and evolves them toward scenarios that likely reveal navigation faults.
  Each generated scenario would be translated into a Player/Stage simulation, where the robot attempts to deliver an order from the kitchen to a target table.
  To discover failures in a specific environment, we could define a new scenario generator based on the FloorPlan DSL\,\cite{parra2023iros}.
  This would enable the generation of more realistic environment descriptions; allow finer-grained control over the environment generation (e.g., restaurant layout or poses and sizes of tables/chairs); and enable running the scenarios in ROS/Gazebo.
  During each run, metrics (e.g., path completion, collision events, travel time) would be recorded as ROS logs.
  Scenarios where the robot gets blocked, collides, or violates safety constraints (e.g., exceeding 0.8~m/s) receive higher fitness scores, guiding the generator toward critical conditions, such as narrow aisles, partially blocked paths, or dense human traffic.
  The resulting scenarios could be replayed to verify improvements after navigation or perception updates.
\end{tcolorbox}

\parhead{Property-Based Testing} Property-based testing validates whether a system fulfills formally specified properties or invariants rather than checking individual input/output pairs\,\cite{fink1997property}.

\looseness=-1
In robotics, property-based testing has been applied to verify runtime invariants in ROS-based systems\,\cite{santos2018esec}. Typical properties include constraints such as ensuring that a robot never leaves a designated workspace or that sensor values remain within valid bounds.
\citet{santos2021rose} introduce the HAROS framework, which provides a ROS-specific property-based testing workflow that extracts system models from source code, converts high-level behavioral properties into Hypothesis-driven test generators, and automatically detects counterexample message traces via runtime monitors.
\citet{robert2020ese} present a simulation-based, property-oriented testing approach in which randomized generation of virtual crop fields and an invariant-based oracle are used to expose navigation failures in an agricultural robot across diverse mission scenarios.
\citet{santos2022schema} automatically generate test cases from metric first-order temporal specifications, allowing high-level behavioral descriptions to drive systematic test generation.
Property-based testing has also been used to assess robot correctness through the automatic creation of simulated test scenarios.
\citet{sohail2021ecmr} propose a property-based simulation framework to verify individual and multiple action tasks, automating test generation and evaluation across varying conditions and parameters.
A related technique, metamorphic testing, extends property-based to situations where precise oracles are difficult to specify. \Citet{zhang2019testing} provide a framework to define metamorphic relations, expected relationships across multiple inputs and outputs, to construct oracles for path-planning algorithms.

\begin{tcolorbox}[myexample]\looseness=-1
 We could apply property-based testing to our ROS-based software with the HAROS framework\,\cite{santos2021rose}. 
  Properties would be expressed in the HAROS Property Language and automatically translated into executable test schemas using the hypothesis library. 
  This would allow the verification of properties such as ``the robot's velocity published on \texttt{/cmd\_vel} must never exceed 0.8~m/s'' (cf.~\cref{tab:requirements}, requirement~10) or ``whenever a \texttt{delivery\_started} message is published, a corresponding \texttt{delivery\_completed} or \texttt{delivery\_failed} message must occur'' (cf.~\cref{tab:requirements}, requirement~7). 
  HAROS would parse the project's ROS packages and launch files to build the communication graph, then inject message sequences and monitor topic values during simulation or runtime. 
  The hypothesis engine would systematically vary message timing, order, and content to explore different system behaviors and reveal counterexamples. 
\end{tcolorbox}

\subsection{White-Box Testing}
\looseness=-1
White-box testing, also referred to as structure-based testing, uses knowledge of the internal structure, or implementation, for deriving test cases by analyzing the source code or other internal representations\,\cite{iso29119pt4}. 
White-box testing is primarily used for unit-level or component-level testing, or low-level integration testing\,\cite{hass2014guide}. 
Testing standards\,\cite{iso29119pt4,istqb2018} specify withe-box testing in terms of coverage metrics, but they do not describe how concrete test inputs are generated.
In practice, test inputs are typically derived either manually by engineers using code inspection and domain knowledge or automatically, using techniques such as symbolic execution, constraint solving, and random or guided input generation. 
In robotics, white-box testing can be used to verify the correctness of motion planning algorithms by ensuring all control paths are exercised\,\cite{kim2005rrt}, or to detect distributed errors in modular robotic systems by leveraging internal module logic\,\cite{de2008distributed}.
\looseness=-1
\textit{Statement Testing}
derives test cases so that every executable statement in the considered part of the source code is executed at least once\,\cite{iso29119pt4}.
The code is analyzed to identify inputs that achieve full statement coverage, which can also reveal dead or unreachable code when certain statements cannot be exercised\,\cite{iso29119pt4,hass2014guide}.
We did not find any papers reporting on statement testing in robotics.





\looseness=-1
\textit{Branch Testing} creates test suites that execute every branch of the control flow graph\,\cite{iso29119pt4,hass2014guide}. 
To this end, the control flow graph is analyzed to identify inputs that trigger each branch in the control flow, potentially revealing unreachable branches.
We did not find any papers reporting on branch testing in robotics.





\looseness=-1
\textit{Decision Testing} creates test cases that execute every decision in the considered part of the source code with all possible outcomes (e.g., true and false) at least once\,\cite{iso29119pt4,hass2014guide}. A decision is any point where control flow can take alternative paths, such as conditions in if, while, or case statements. Inputs are identified that trigger each possible decision outcome. The focus is on exercising each decision outcome at least once. Therefore, the resulting test suite subsumes branch testing.
We did not find any papers reporting on decision testing in robotics.






\looseness=-1
\textit{Branch Condition Testing}
extends decision testing with a focus on the evaluation of individual atomic conditions rather than only the overall decision outcome\,\cite{iso29119pt4,hass2014guide}.
For example, for the expression \texttt{if (A or (B and C))}, decision testing considers only whether the entire expression evaluates to true or false, whereas branch condition testing requires each atomic condition (i.e., \texttt{A}, \texttt{B}, and \texttt{C}) to independently take both true and false values at least once.
Branch condition combination testing further requires the execution of all possible combinations of atomic condition values, resulting in $2^N$ test cases for $N$ conditions\,\cite{iso29119pt4,hass2014guide}. Modified condition decision coverage refines this by finding all combinations of atomic conditions within a decision that independently affect its outcome\,\cite{iso29119pt4,hass2014guide}.
Test cases are derived based on the atomic conditions.
We did not find any papers reporting on branch condition-based testing technique in robotics.

\textit{Data Flow Testing}
focuses on how variables are defined, used, and deleted within the source code. 
Test cases are derived by analyzing the control flow to identify definitions (value assignments) and uses (value is read, used either in computations or in decisions). For each definition-use (DU) pair, inputs are selected that execute the DU-path\,\cite{iso29119pt4}.
The granularity of the paths can change, either checking that every definition is reached by at least one test (all-definitions), that definitions flow into computational uses (all-c-uses), into predicate uses (all-p-uses), or that both types of uses (all-uses). All-du-paths is the most rigorous variant and checks every possible path of each definition to its uses\,\cite{iso29119pt4}.
We did not find any papers reporting on data flow testing in robotics.






\subsection{Experience-Based Testing}
This test design technique relies on the developer's knowledge instead of systematically deriving test cases from models, code, or other specifications. The developer identifies relevant aspects of the system and creates test cases, that is, test cases are derived in an ad hoc manner\,\cite{iso29119pt4,hass2014guide}.

\looseness=-1
\textit{Error-Guessing} derives test cases from knowledge of error types and sources of likely errors in the system. Test cases can be derived from experiences, error taxonomies or lists, bug or incident trackers, or other knowledge bases\,\cite{iso29119pt4,hass2014guide}.
We did not find papers reporting error-guessing in robotics.

\looseness=-1
\textit{Exploratory Testing} enables developers to design and adapt tests spontaneously to discover ``hidden properties'' of the system that may lead to failures\,\cite{iso29119pt1}.
Test design relies on the developers' domain knowledge, heuristics about the system's behavior, and insights from previous test results.
In robotics, exploratory experiments have been proposed as an alternative to controlled testing, as unpredictable real-world environments make predefined outcomes difficult to specify.
\citet{amigoni2016explorative} introduces exploratory experiments in autonomous robotics since conducting traditional controlled experiments can be challenging due to unpredictable environments. 






\subsection{Fuzz Testing}
Fuzz testing\,\cite{miller1990empirical}, or fuzzing, is a test case generation technique based on large volumes of random or pseudo-random inputs\,\cite{iso29119pt1}.
Unlike systematic techniques, fuzzing does not require an explicit specification of expected functional behavior. Instead, it relies on observing program reactions to stressed input conditions, such program crashes detection, timeouts, memory safety violations, or other sanitizer-reported errors.
Depending on the available system knowledge and instrumentation, fuzzing can be applied in black-box, white-box, or grey-box testing, with feedback mechanisms (e.g., code coverage) often used to guide input generation and improve fault discovery.

Fuzzing frameworks for robotics, such as ROZZ\,\cite{xie20222icra} and RoboFuzz\,\cite{kim2022robofuzz}, extend traditional fuzzing by supporting multi-dimensional input generation, distributed node coverage, and temporal mutation strategies. These frameworks can target both low-level software faults (e.g., memory errors, crashes) and higher-level semantic or cyber-physical bugs (e.g., violations of domain-specific invariants or real-world constraints).
RROSFuzz tests the consistency and robustness of robotic software by generating and mutating inputs for different versions or implementations of ROS packages and comparing their behavior in RViz\,\cite{wang2022toward}. PHYS-FUZZ incorporates physical semantics such as robot size, trajectory, and time-to-impact into input generation, effectively identifying collision-prone or hazardous navigation behaviors in simulation environments like Gazebo\,\cite{woodlief2021fuzzing}.

\looseness=-1
Several works focus on uncovering faults in control and behavior logic. RVFuzzer uses control-guided fuzzing for robotic vehicle software, mutating control parameters and using control instability as feedback to reveal input validation bugs leading to unsafe or unstable motion\,\cite{kim2019rvfuzzer}. Similarly, \citet{delgado2021fuzz} apply grammar-based fuzzing to robotic behaviors in the SMACH framework for ROS, generating structured state-machine inputs to identify coding, logic, and configuration errors. 
MDPFuzz focuses on how decision policies learned by models based on Markov Decision Processes, such as reinforcement learning and planning-based controllers, behave under unexpected or adversarial conditions\,\cite{pang2022mdpfuzz}. It mutates initial environment states and observes whether resulting trajectories lead to unsafe or abnormal outcomes.

In robotics, fuzz testing is particularly useful for identifying robustness and security issues in complex, distributed systems.
\citet{WANG2021101838} use directed fuzzing with RoboFuzz to inject rational but harmful sensor values (e.g., falsified distance readings) in line with the robot's state and environment, revealing vulnerabilities in control logic that can lead to unsafe behavior.

\subsection{Field-Based Testing}\label{sec:sub:field-based-testing}
\looseness=-1
Field-based testing is a test design technique in which test cases are executed in, or derived from, the production environment.
Although not defined in existing testing standards, it plays a central role in robotics, where realistic environmental conditions and interactions cannot be fully reproduced in laboratory settings.
Field tests are typically performed at the end-to-end or acceptance level to assess both functional and non-functional requirements and can be categorized as ex-vivo, executed in the development environment (in-house) using real-world data to replay field scenarios, or in-vivo, executed directly in the production environment.
Tests may run offline on a separate system instance, reducing operational risks, or online on the deployed system itself.
Ex-vivo and offline testing primarily target functional faults, whereas online testing also addresses non-functional properties such as performance, security, and robustness\,\cite{bertolino2021survey}.
In robotics, field-based testing serves different goals, including exploratory field-testing, which aims to uncover unexpected behaviors and corner cases in a real-world context, and endurance field-testing, which evaluates the robot's long-term performance and robustness under sustained load in the production environment, usually performed at the integration or system level\,\cite{ortega2022iros, luperto2021my}.
Despite being resource intensive and requiring a fully built system, field-based testing yields insights not capturable through laboratory or simulation-based testing\,\cite{shin2022hands, gasteiger2022participatory}.
Field-based testing typically requires instrumenting the software and building it for that purpose, which has been studied for ROS-based systems\,\cite{caldas2024runtime}.

\looseness=-1
In the literature, field-based testing is often conducted in an ad hoc manner to evaluate system performance, reliability, or component behavior in real-world conditions\,\cite{erdos2008uav,yu2007development, li2019uav}.
Numerous studies report deployments and evaluations of robots in the field, underscoring the importance of field-based testing for acceptance testing and safety\,\cite{kuehn2017system, jiang2023design, fink2009multi}.
Considering different types of robots such as UAVs\,\cite{michael2010grasp,suarez2021experimental,cwikakala2019testing,verdu2020experimental,garcia2009implementation,montgomery2006jet}, UGVs\,\cite{sadek2022depth,ahuja2009test,gregory2022active,chainarong2022development}, or humanoid robots\,\cite{kuehn2017system,banerjee2018integrated,watanabe2022hardware}.
Field-based tests can be applied to different domains, such as agriculture\,\cite{xiong2019development,xiong2020autonomous,birrell2020field,zhou2022design,pak2022field}, heavy duty\,\cite{gucunski2017rabit,Tanaka17012020,paul2010robotic}, healthcare\,\cite{mauch2017service,lundberg2022robotics,luperto2021my}, or space\,\cite{barfoot2011field,wettergreen2010design,lamboley1995marsokhod}.
Often not only testing a single robot but also multi-robot systems in cooperative tasks\,\cite{conte2018development,fink2009multi,park2023multiple,pook1999test,leonard2010coordinated,kapetanovic2022heterogeneous}.

\section{Challenges and Gaps}
Despite many techniques, testing of robots still faces many challenges. Some of the surveyed papers explicitly mention challenges, which we summarize below, based on five categories we identified.

\parhead{Simulation-Reality Gap}
As briefly discussed in \cref{sub:sec:sim-based}, the simulation-reality gap describes the difference in fidelity of simulations and real-world\,\cite{aljalbout2025reality}.  
While simulations play a crucial role in robotics, creating a simulation providing the same fidelity as the real world is costly and exhausting and often infeasible\,\cite{timperley2018icst}.
Robotic simulations struggle to accurately model real-world physics, sensor inaccuracies, mechanical wear, and environmental complexity\,\cite{sunspiral2012development}. Factors such as friction, deformation, and imperfect sensor calibration introduce discrepancies between simulated and real-world performance\,\cite{afzal2021icst}. 
Furthermore, modeling the behavior of moving objects, especially humans (e.g., pedestrians), is often infeasible in simulations\,\cite{kaur2022simulators, kaneshige2021overcome,bostelman2012standard}.

\looseness=-1
Substantial efforts are made in the robotics community to use simulations effectively, supported by our results in \cref{sub:sec:sim-based}. Many papers address the issue of how to build a simulation environment with a sufficient level of fidelity. However, standard simulation for different robot types is still missing.

\parhead{Lack of Standards}
\looseness=-1
Missing standards, unified testing processes and a variety of available middleware, such as ROS\,\cite{quigley2009icra}, YARP\,\cite{metta2006yarp}, and SORA\,\cite{fluckiger2014service}, can lead to ad hoc testing strategies. The broad application scope of robotic systems, from human-robot interaction to autonomous delivery drones, makes it especially complex to define standards\,\cite{poncela2013framework, bostelman2013development,bostelman2012standard, marrella2022methodology} or standard test courses for a uniform validation of the system\,\cite{jacoff2001reference, jimenez2013testbeds}. Therefore, results across studies are hard to compare directly with each other, limiting overall reproducibility.
To overcome some of these challenges, \citet{caldas2024runtime} propose guidelines to help developers and quality assurance teams when developing and testing robots, especially tailored to field-testing. 

The lack of standards often occurs along with a lack of coverage for both requirements and systems (e.g., code coverage).
While standards provide guidelines and structure, coverage must be defined to build trust in the system.
Especially, for robotics systems where the input space is effectively infinite\,\cite{afzal2020icst,micskei2012amsta}, systematically covering corner cases is important.
Unlike traditional software systems, which are often deterministic and test cases can be well-defined, robotics operates in dynamic and often unpredictable environments. Additionally, sensor noise, actuator variability, and human-robot interaction, make identifying corner cases complex and often unfeasible\,\cite{afzal2020icst, micskei2012amsta, totev2021testing}. For example, it remains difficult to determine the correct behavior for pedestrians\,\cite{kaneshige2021overcome, kaur2022simulators, bostelman2012standard, marrella2022methodology}.
An unsystematic testing procedure often results in incomplete or absent coverage criteria and relies on developers' domain knowledge, thereby failing to systematically cover corner cases\,\cite{micskei2012amsta,totev2021testing}.

In robotics, test design techniques, such as scenario-based testing, address these issues by deriving test cases from specific scenes. Especially DSLs and modular scenario descriptions provide a basis to describe complex scenes systematically. 
While they are first steps to systematic coverage, they can still generate unrealistic scenarios\,\cite{micskei2012amsta,bostelman2012standard,bostelman2013development,timperley2018icst}.
	Property-based testing addresses the challenge of the infinite input space by defining properties that must always hold. 
	This allows to abstract details for each test case and focus on necessary invariants or safety requirements of the system.

\looseness=-1
\parhead{Test Automation}
While standards and coverage are important to structure and guide testing, test automation can accelerate the process and does not necessarily need an underlying standard.
Test automation has two main steps, (1) automatically creating meaningful test cases, and (2) developing a pipeline for automatic execution and evaluation.
For traditional software development, continuous integration and deployment pipelines are well established, however, for developing robotic systems, these pipelines still lack behind\,\cite{kaur2022simulators,micskei2012amsta,afzal2021icst}.
While using simulations for automation can be powerful, it is challenging\,\cite{afzal2021icst}. Often, automation is tailored to specific simulators and environment representations, making reuse across multiple projects difficult.
Furthermore, robotics relies heavily on field-based testing, making test automation nearly impossible, since testing in physical environments is difficult to automate and only a very limited number of scenarios can be executed\,\cite{nybacka2009opportunities,jacoff2001reference, poncela2013framework}.

\parhead{Reproducibility and Scalability}
Reproducibility and scalability are general cross-domain problems across. Due to the hardware nondeterminism, environmental variability, and human involvement in robotic systems, they are particular challenging in this domain. No real-world conditions are alike\,\cite{nybacka2009opportunities,poncela2013framework}, hardware variation and sensor noise introduce nondeterminism even under nominally identical inputs\,\cite{afzal2021icst,totev2021testing}. Furthermore, simulators are not fully deterministic, producing different results across different machines\,\cite{timperley2018icst}. In summary, the reproducibility is not only limited across different projects, but also of two versions of the same robot, as small deviations in hardware can lead to different behavior.
On the other hand, scalability is limited by physical environments, which naturally do not scale well, and only a limited number of test cases can be executed per day\,\cite{nybacka2009opportunities,poncela2013framework,jacoff2001reference}. Simulations can compensate for this problem, but running large numbers of simulation tests is computationally expensive and often unfeasible\,\cite{kaur2022simulators}. Furthermore, more test design techniques, such as scenario-based testing, do not scale linearly and multi-robot systems can create combinatorial explosions of coordination scenarios\,\cite{poncela2013framework,nybacka2009opportunities,micskei2012amsta}.

\looseness=-1
\parhead{Oracle Problem}
Determining the expected outcome of a test---a.k.a. the oracle problem---is a general challenge in software testing, and particularly difficult in robotics\,\cite{barr2014oracle}. 
While robotics inherits this problem from traditional software testing, such as incomplete specifications and ambiguous requirements, real-world uncertainties, physical interactions, and hardware variability only complicate it\,\cite{afzal2020icst}.
This makes it difficult to predict the correct system response, as the distinction between correct, acceptable, and incorrect system behavior blurs\,\cite{afzal2020icst, aliabadi2020ikt}.
Often multiple responses can be correct, solving the problem in different ways.
To tackle the oracle problem, robotics testing heavily relies on MBT, scenario testing, and property-based testing, which also provide the ground truth. However, developing a model is still a complex, challenging task.
\section{Discussion}
\looseness=-1
We learned that system-level testing (e.g., end-to-end and acceptance testing) conducted in late development stages is dominant in the robotics testing literature. 
Specifically, black-box techniques at the system level, such as scenario-based and property-based testing, are most prevalent.
This trend is also reflected in the widespread use of simulation- and field-based testing, both of which naturally align with system-level evaluation or acceptance testing, as they exercise the robot as an integrated whole rather than isolated components.
Testing in earlier development stages, such as unit and integration testing, is underrepresented in the literature.
This heavy reliance on late-stage system-level testing can be problematic, as it complicates fault localization.
When failures are detected only at the system level, it often remains unclear whether the root cause lies in an individual component, an interaction between components, or the overall system architecture.
This increases debugging efforts and raises the cost of fault resolution.
Future research should \textbf{bring systematic testing activities into earlier development stages}, in line with established software testing standards.
In particular, tailored white-box testing techniques for, e.g., control, perception, are needed to enable systematic coverage at unit and integration testing levels.
Integrating such techniques into robotics middleware and development frameworks would support continuous testing throughout development.


\looseness=-1
We observed that more emphasis is on non-functional testing, such as robustness and performance, rather than functional correctness.
The distinction between functional and non-functional testing often blurs in robotics, as properties such as safety, robustness, and performance are treated as functional requirements rather than ``could-have'' quality attributes.
This observation suggests that traditional testing taxonomies may require adaptation when applied to robotics.
Treating robustness and safety as non-functional properties risks under-specifying expected system behavior, which in turn hampers systematic test design and assessment.
Future research should work \textbf{toward clearer specification and classification of robotic system requirements}, enabling more systematic test design that explicitly links functional expectations with measurable system qualities.

\looseness=-1
Our survey further revealed that testing processes in robotics are often ad hoc and experiment-driven. 
Test cases are derived from usage scenarios or domain knowledge instead of systematic methods.
Corner cases are frequently identified based on developers' experience and the robot's operational domain.
Such practices limit the repeatability and reliability of testing, which can affect the trustworthiness of robots.
Without systematic coverage or well-defined test criteria, it is difficult to assess whether a system has been sufficiently validated before deployment.
Moreover, the absence of systematic testing and standards hinders reproducibility and comparability across studies, which can negatively impact the credibility of experimental results in the literature.
Future research should \textbf{define lightweight, systematic, and resource aware testing-processes that align with established software testing standards} while accounting for constraints in robotics, such as limited hardware availability, restricted testing time, and limited computational resources\,\cite{gotlieb2020testing}.
Standardization in both testing processes and reporting practices would significantly strengthen the reliability and comparability of results within the robotics community.

\looseness=-1
Finally, addressing the simulation-to-reality gap remains a critical challenge\,\cite{aljalbout2025reality}. 
While simulation-based testing enables scalability and early feedback, field-based testing is essential for exposing real-world uncertainties and unmodeled effects.
\textbf{Advancing hybrid testing strategies that systematically combine simulation-based and field-based testing} could improve the confidence in system behavior under real operating conditions.
\textbf{Further automation in test generation, execution, and evaluation}, supported by unified testing processes, would improve scalability, reduce manual effort, and accelerate testing cycles.
\section{Related Work}


\looseness=-1
Existing reviews of the literature only address specific aspects of testing robotics systems, such as formal verification, simulation tools, or safety assurance.
%
\citet{luckcuck2020csur} conduct a literature survey on formal specification and verifications for autonomous robotic systems. They identify the unique challenges of these systems and evaluate the current state of the art. 
Santos et al.~\cite{santos2021arxiv} conduct a mapping study of software engineering for robotics, identifying key concepts and practices. Their work highlights the need for more structured approaches to software engineering in robotics, especially for testing methodologies, as they are underrepresented. However, they do not provide a structured overview of testing robotic systems and focus more on general software engineering practices, instead of testing.
\citet{song2021wain} conduct an industrial and academic study to identify concepts of robot testing, bridging the gap between academic research and industrial need.
%
\citet{garousi2018testing} provide a systematic literature mapping of the state-of-the-art practices in embedded software systems. They reviewed 312 primary studies, categorizing them by contributions across testing techniques, tools, and identified trends in industry and academia. The study organizes the research and guides practitioners to understand existing methods, avoid redundancy in their testing, and identifies open challenges for testing embedded systems.
\citet{araujo2022tse} conduct the first large-scale systematic literature review of validation and verification techniques for robotic systems, analyzing 195 studies. They find that research is dominated by formal models (e.g., temporal logic, state machines) and generic adequacy measures, with tool use centered on model checkers and simulators. The study highlights gaps in domain-specific metrics, consolidated benchmarks, and industrial evidence. 
In contrast, our study focuses on mapping robotics testing literature to established software testing standards and taxonomies, offering a perspective that emphasizes alignment and coverage rather than technique-centric analysis---enabling us to uncover how established software testing concepts are adopted or missing in robotic software testing.
\section{Conclusion}%
\label{sec:conclusion}
\looseness=-1
We presented the state-of-the-art in robotics testing in relation to the software testing theory. By mapping the theory to the reality in robotics, we identified relevant techniques from the software testing body-of-knowledge, allowing us to uncover gaps in research and practice. 
We also described challenges reported in the literature and discussed how they are currently addressed.
Our analysis revealed that robot testing is often performed at the system level, focusing on end-to-end or acceptance tests. Unit and integration tests are rarely found, and black-box techniques dominate, with limited application of white-box techniques. The testing processes are often ad hoc and heavily rely on expert experience rather than standardized procedures. These findings indicate that robotics lacks systematic and repeatable testing practices. To advance them, future work should prioritize the establishment of standards, the definition of structured testing processes, and the reduction of the simulation-to-reality gap, which remains a persistent challenge for ensuring reliable and transferable results.

\begin{acks}
Partly supported by the EU Horizon 2020 project SESAME (no. 101017258) and the German Federal Ministry of Research, Technology and Space (BMFTR) under the Robotics Institute Germany (RIG).
\end{acks}








\bibliographystyle{ACM-Reference-Format}
\bibliography{definitions,merged-no-doi}

@STRING{icra    = {ICRA} }

@STRING{icst    = {ICST } }

@STRING{iros    = {IROS} }

@STRING{iv      = {IV} }

@STRING{jfr     = {JFR} }

@STRING{jirs    = {J. of Intell. \& Robot. Syst.} }

@STRING{ral        = "{IEEE} Robot. Autom. Lett."}

@STRING{ram         = "RAM"}

@STRING{ras     = {Robot. \& Auton. Syst.} }

@STRING{roman     = {RO-MAN} }

@STRING{rssbook = {RSS} }

@STRING{springer= {Springer} }

@STRING{tro         = "{IEEE} Trans. Robot."}

@STRING{ijrr     = {IJRR} }

@STRING{esec     = {ESEC/FSE}}

@STRING{dessert     = {DESSERT} }

@STRING{ecmr     = {ECMR} }

@STRING{qrs     = {QRS} }

@STRING{irc     = {IRC} }

@STRING{wain    = {WAIN}}

@STRING{models    = {MODELS}}

@STRING{sii    = {Int.~Symp.~on Syst. Integration}}

@STRING{ijra    = {IJIRA}}

@STRING{tse         = "TSE"}

@STRING{tii         = "{IEEE} Trans. Ind. Informat."}

@STRING{tc         = "{IEEE} Trans. Comput."}

@STRING{ssrr    = {SSRR}}

@STRING{hri    = {HRI}}

@STRING{mfi    = {MFI}}

@STRING{aitest    = {AITest}}

@STRING{aqtr    = {AQTR}}

@STRING{isr    = {ISR}}

@STRING{ast    = {AST}}

@STRING{jirs    = {J. Intell. Robot. Syst.} }

@STRING{issta    = {ISSTA}}

@STRING{sose    = {SoSE}}

@STRING{ubicomp    = {UbiComp}}

@STRING{ijcasa    = {Int. J.~of Adv. Comp. Sci. and Applications} }

@STRING{ijsea    = {JSEA} }

@STRING{edcc    = {EDCC}}

@STRING{ijars    = {IJARS} }

@STRING{ijsr    = {Int. J. Soc. Robot.} }

@STRING{ijamt    = {Int. J.~of Adv. Manufacturing Technology} }

@STRING{icsoftea    = {ICSOFT-EA}}

@STRING{esec    = {ESEC/FSE}}

@STRING{icsa    = {ICSA}}

@STRING{ist    = {IST} }

@STRING{mesas    = {MESAS}}

@STRING{simpar    = {SIMPAR}}

@STRING{ki    = {KI}}

@STRING{sfm    = {SFM}}

@STRING{phmconf    = {PHM Conf}}

@STRING{permis    = {PerMIS}}

@STRING{acra    = {ACRA}}

@STRING{icar    = {ICAR}}

@STRING{idetc    = {IDETC/CIE}}

@STRING{icinco    = {ICINCO}}

@STRING{fose    = {FOSE}}

@STRING{re    = {RE}}

@STRING{incet    = {INCET}}

@STRING{access    = {IEEE Access}}

@STRING{arxiv    = {arXiv}}

@STRING{emsoft    = {EMSOFT}}

@STRING{morse    = {MORSE}}

@STRING{kes-amsta    = {KES-AMSTA}}

@STRING{case    = {CASE}}

@STRING{cec    = {CEC}}

@STRING{entcs    = {Electronic Notes in Theoretical Comp. Sci.}}

@STRING{tosem    = {TOSEM}}

@STRING{civemsa    = {CIVEMSA}}

@STRING{rose    = {RoSE}}

@STRING{icse-companion    = {ICSE Companion Proc.}}

@STRING{indin    = {INDIN}}

@STRING{spce    = {SPCE}}

@STRING{plans    = {PLANS}}

@STRING{vtc    = {VTC}}

@STRING{iccve    = {ICCVE}}

@STRING{prdc    = {PRDC}}

@STRING{acmsen    = {Softw. Eng. Notes}}

@STRING{acmcs    = {CSUR}}

@STRING{a-test    = {A-TEST}}

@STRING{ictke    = {ICT\&KE}}

@STRING{dcoss    = {DCOSS}}

@STRING{scp    = {Sci.~of Comp. Programming}}

@STRING{compag    = {Comput. Electron. Agric.}}

@STRING{icites    = {ICITES}}

@STRING{metrocad    = {MetroCAD}}

@STRING{icabcd    = {icABCD}}

@STRING{enase    = {ENASE}}

@STRING{eptcs    = {Electron. Proc. Theor. Comput. Sci.}}

@STRING{rcim    = {Robotics and Computer-Integrated Manufacturing}}

@STRING{daaam    = {DAAAM Int. Symp.}}

@STRING{airpharo    = {AIRPHARO}}

@STRING{ikt    = {IKT}}

@STRING{sqj    = {Softw. Quality J.}}

@STRING{ihsed    = {IHSED}}

@STRING{ser    = {Softw. Eng. for Robotics}}

@STRING{jrie    = {J. Reliab. Intell. Environ.}}

@STRING{sice    = {SICE}}

@STRING{ese    = {EMSE}}

@STRING{simultech    = {SIMULTECH}}

@STRING{red-uas    = {RED UAS}}

@STRING{usenix    = {USENIX Security Symp.}}

@STRING{taai    = {TAAI}}

@STRING{ssms    = {SSMS}}

@STRING{tools    = {TOOLS}}

@STRING{mases    = {MASES}}

@STRING{sdl    = {SDL}}

@STRING{aemusp    = {AHFE}}

@STRING{mbr    = {MBR}}

@STRING{taros    = {TAROS}}

@STRING{wpc    = {Wireless Personal Communications}}

@STRING{frontiers-ai    = {Front. Robot. AI}}

@STRING{jsep    = {J.~of Software: Evolution and Process}}

@STRING{applied-sci    = {Appl. Sci.}}

@STRING{cmc    = {CMC}}

@STRING{jhri    = {J. Hum.-Robot Interact.}}

@STRING{sensors    = {Sensors}}

@STRING{robotics    = {Robotics}}

@STRING{adv-robotics    = {AR}}

@STRING{auton-robots    = {Auton. Robots}}

@STRING{msc    = {Mechatron. Syst. Control.}}

@STRING{jsa    = {J.~of Systems Architecture}}

@STRING{electronics    = {Electronics}}

@STRING{ea-chi    = {CHI}}

@STRING{wsc    = {Winter Simulation Conf.}}

@STRING{is    = {IEEE Intell. Syst.}}

@STRING{mmif    = {Multisensor, Multisource Information Fusion: Architectures, Algorithms, and Applications}}

@STRING{spie-ust    = {SPIE Unmanned Syst. Technology}}

@STRING{spie-set    = {SPIE Space Exploration Technologies}}

@STRING{spie-ugvt    = {SPIE Unmanned Ground Vehicle Technology}}

@STRING{spie-mr    = {SPIE Mobile Robots}}

@STRING{jrobotics    = {J.~of Robotics}}

@STRING{ems    = {ESIAM}}

@STRING{modevva    = {MoDeVVa@MoDELS}}

@STRING{icra-oss    = {ICRA Workshop Open Source Softw.}}

@STRING{ieee-r5    = {IEEE Reg. 5 Conf.}}

@inproceedings{nguyen2025,
  title        = {Automated {{Behaviour-Driven Acceptance Testing}} of {{Robotic Systems}}},
  author       = {Nguyen, Minh and Wrede, Sebastian and Hochgeschwender, Nico},
  year         = 2025,
  booktitle    = iros
}

@Article{santos2024frontiers,
  author  = {dos Santos, Marcela G. and Hall{\'{e}}, Sylvain and Petrillo, Fabio and Gu{\'{e}}h{\'{e}}neuc, Yann-Ga{\"{e}}l},
  journal = frontiers-ai,
  title   = {AAT4IRS: automated acceptance testing for industrial robotic systems},
  year    = {2024},
  volume  = {11},
}

@Article{ortega2024frontiers,
  author  = {Ortega, Argentina and Parra, Samuel and Schneider, Sven and Hochgeschwender, Nico},
  journal = frontiers-ai,
  title   = {Composable and Executable Scenarios for Simulation-Based Testing of Mobile Robots},
  year    = {2024},
  volume  = {11},
}

@incollection{weinstein2024equivalent,
  title        = {Equivalent Environments and Covering Spaces for Robots},
  author       = {Vadim K. Weinstein and Steven M. LaValle},
  year         = 2024,
  booktitle    = ems
}

@inproceedings{cheng2023simulation,
  title        = {Simulation and Verification of Underwater Robot Target Grasping based on PID Algorithm},
  author       = {Cheng, Peng},
  year         = 2023,
  booktitle    = incet
}

@inproceedings{golchinfar2023let,
  title        = {Let me Be your Service Robot: Exploring Early User Experiences of Human-Robot Collaboration for Service Domains},
  author       = {Golchinfar, David and Vaziri, Daryoush and Hennekeuser, Darius and Stevens, Gunnar and Schreiber, Dirk},
  year         = 2023,
  booktitle    = roman
}

@Article{humeniuk2023ambiegen,
  author  = {Humeniuk, Dmytro and Khomh, Foutse and Antoniol, Giuliano},
  journal = scp,
  title   = {Ambiegen: A search-based framework for autonomous systems testing},
  year    = {2023},
  volume  = {230},
}

@article{jiang2023design,
  title        = {Design and analysis of a wall-climbing robot for water wall inspection of thermal power plants},
  author       = {Jiang, Ze and Chen, Bo and Ju, Zhongjin and Li, Yichao and Xu, Yundou and Zhao, Yongsheng},
  year         = 2023,
  journal      = jfr
}

@Article{jimenez2023automated,
  author  = {Jim{\'{e}}nez-Ram{\'{i}}rez, Andres and Chac{\'{o}}n-Montero, Jes{\'{u}}s and Wojdynsky, Tomasz and Gonz{\'{a}}lez Enr{\'{i}}quez, Jos{\'{e}}},
  journal = jsep,
  title   = {Automated testing in robotic process automation projects},
  year    = {2023},
  number  = {3},
  volume  = {35},
}

@Article{kim2023development,
  author  = {Kim, Shinhyung and Philip, Opayemi Alaba and Tullu, Abera and Jung, Sunghun},
  journal = access,
  title   = {Development and Verification of a ROS-Based Multi-DOF Flight Test System for Unmanned Aerial Vehicles},
  year    = {2023},
  volume  = {11},
}

@Article{kluck2023empirical,
  author  = {Kl{\"u}ck, Florian and Li, Yihao and Tao, Jianbo and Wotawa, Franz},
  journal = ist,
  title   = {An empirical comparison of combinatorial testing and search-based testing in the context of automated and autonomous driving systems},
  year    = {2023},
  volume  = {160},
}

@Article{nunez2023implementation,
  author  = {Manuel N{\'{u}}{\~{n}}ez and Robert M. Hierons and Raluca Lefticaru},
  journal = ras,
  title   = {Implementation relations and testing for cyclic systems: Adding probabilities},
  year    = {2023},
  volume  = {165},
}

@article{neri2023experimental,
  title        = {Experimental evaluation of collision avoidance techniques for collaborative robots},
  author       = {Neri, Federico and Forlini, Matteo and Scoccia, Cecilia and Palmieri, Giacomo and Callegari, Massimo},
  year         = 2023,
  journal      = applied-sci,
  VOLUME = {13},
  NUMBER = {5},
  ARTICLE-NUMBER = {2944},
}

@Article{park2023multiple,
  author  = {Park, Jeonghong and Kang, Minju and Lee, Yeongjun and Jung, Jongdae and Choi, Hyun-Taek and Choi, Jinwoo},
  journal = access,
  title   = {Multiple autonomous surface vehicles for autonomous cooperative navigation tasks in a marine environment: Development and preliminary field tests},
  year    = {2023},
  volume  = {11},
}

@inproceedings{parra2023iros,
  title        = {A Thousand Worlds: {{Scenery}} Specification and Generation for Simulation-Based Testing of Mobile Robot Navigation Stacks},
  author       = {Parra, Samuel and Ortega, Argentina and Schneider, Sven and Hochgeschwender, Nico},
  year         = 2023,
  booktitle    = iros
}

@Article{sim2023sminer,
  author  = {Sim, Kyungmin and Yi, Jeong Hyun and Cho, Haehyun},
  journal = cmc,
  title   = {SMINER: Detecting Unrestricted and Misimplemented Behaviors of Software Systems Based on Unit Test Cases.},
  year    = {2023},
  number  = {2},
  volume  = {75},
}

@inproceedings{sohail2023case,
  title        = {Automated Testing of Standard Conformance for Robots},
  author       = {Sohail, Salman Omar and Schneider, Sven and Hochgeschwender, Nico},
  year         = 2023,
  booktitle    = case
}

@Article{mashrur2023assistive,
  author  = {Tanzim Mashrur and Zeyad Ghulam and Gregory French and Hussein A Abdullah},
  journal = ijars,
  title   = {Assistive feeding robot for upper limb impairment—Testing and validation},
  year    = {2023},
  number  = {4},
  volume  = {20},
}

@book{dslbook,
  title        = {Domain-Specific Languages},
  author       = {W{\k{a}}sowski, Andrzej and Berger, Thorsten},
  publisher    = {Springer},
  year         = 2023
}

@Article{araujo2022tse,
  author  = {Araujo, Hugo and Mousavi, Mohammad Reza and Varshosaz, Mahsa},
  journal = {TOSEM},
  title   = {Testing, Validation, and Verification of Robotic and Autonomous Systems: A Systematic Review},
  year    = {2023},
  number  = {2},
  volume  = {32},
}

@Article{bedard2022iral,
  author  = {B{\'{e}}dard, Christophe and L{\"{u}}tkebohle, Ingo and Dagenais, Michel},
  journal = {RA-L},
  title   = {ros2\_tracing: Multipurpose Low-Overhead Framework for Real-Time Tracing of ROS 2},
  year    = {2022},
  number  = {3},
  volume  = {7},
}

@inproceedings{chainarong2022development,
  title        = {Development of an Obstacle Avoidance System for Autonomous Material Handling Vehicles},
  author       = {Chainarong, S. and Yaovaja, K.},
  year         = 2022,
  booktitle    = ictke
}

@inproceedings{dey2022synchrosim,
  title        = {SynchroSim: An Integrated Co-simulation Middleware for Heterogeneous Multi-robot System},
  author       = {Dey, Emon and Hossain, Jumman and Roy, Nirmalya and Busart, Carl},
  year         = 2022,
  booktitle    = dcoss
}

@article{gasteiger2022participatory,
  author = {Gasteiger, N. and Ahn, H. S. and Lee, C. and Lim, J. and MacDonald, B. A. and Kim, G. H. and Broadbent, E.},
  title = {Participatory Design, Development, and Testing of Assistive Health Robots with Older Adults: An International Four-year Project},
  year = {2022},
  volume = {11},
  number = {4},
  journal = jhri,
}

@Article{gleirscher2022verified,
  author  = {Gleirscher, Mario and Calinescu, Radu and Douthwaite, James and Lesage, Benjamin and Paterson, Colin and Aitken, Jonathan and Alexander, Rob and Law, James},
  journal = scp,
  title   = {Verified synthesis of optimal safety controllers for human-robot collaboration},
  year    = {2022},
  volume  = {218},
}

@inproceedings{gregory2022active,
  title        = {Active Learning for Testing and Evaluation in Field Robotics: A Case Study in Autonomous, Off-Road Navigation},
  author       = {Gregory, Jason M. and Sahu, Daniel and Lancaster, Eli and Sanchez, Felix and Rocks, Trevor and Kaukeinen, Brian and Fink, Jonathan and Gupta, Satyandra K.},
  year         = 2022,
  booktitle    = icra
}

@Article{hamann2021scalability,
  author  = {Hamann, Heiko and Reina, Andreagiovanni},
  journal = tc,
  title   = {Scalability in Computing and Robotics},
  year    = {2022},
  number  = {6},
  volume  = {71},
}

@Article{huck2021testing,
  author  = {Huck, Tom P. and Ledermann, Christoph and Kr{\"{o}}ger, Torsten},
  journal = ral,
  title   = {Testing Robot System Safety by Creating Hazardous Human Worker Behavior in Simulation},
  year    = {2022},
  number  = {2},
  volume  = {7},
}

@Article{humeniuk2022search,
  author  = {Humeniuk, Dmytro and Khomh, Foutse and Antoniol, Giuliano},
  journal = ist,
  title   = {A search-based framework for automatic generation of testing environments for cyber--physical systems},
  year    = {2022},
  volume  = {149},
}

@Article{kamandar2022design,
  author  = {Kamandar, Mohammad Reza and Massah, Jafar and Jamzad, Mansour},
  journal = compag,
  title   = {Design and evaluation of hedge trimmer robot},
  year    = {2022},
  volume  = {199},
}

@inproceedings{kang2022behavior,
  title        = {Behavior-Tree Based Scenario Specification and Test Case Generation for Autonomous Driving Simulation},
  author       = {Kang, Shuting and Hao, Haoyu and Dong, Qian and Meng, Lingzhong and Xue, Yunzhi and Wu, Yanjun},
  year         = 2022,
  booktitle    = icites
}

@Article{kapetanovic2022heterogeneous,
  author  = {Kapetanovi{\'{c}}, Nadir and Gori{\v{c}}anec, Jurica and Vatavuk, Ivo and Hrabar, Ivan and Stuhne, Dario and Vasiljevi{\'{c}}, Goran and Kova{\v{c}}i{\'{c}}, Zdenko and Mi{\v{s}}kovi{\'{c}}, Nikola and Antolovi{\'{c}}, Nenad and Ani{\'{c}}, Marina and Kozina, Bernard},
  journal = sensors,
  title   = {Heterogeneous Autonomous Robotic System in Viticulture and Mariculture: Vehicles Development and Systems Integration},
  year    = {2022},
  number  = {8},
  volume  = {22},
}

@inproceedings{kaur2022simulators,
  title        = {Simulators for Mobile Social Robots: State-of-the-Art and Challenges},
  author       = {Kaur, Prabhjot and Liu, Zichuan and Shi, Weisong},
  year         = 2022,
  booktitle    = metrocad
}

@inproceedings{kim2022robofuzz,
  title        = {RoboFuzz: fuzzing robotic systems over robot operating system (ROS) for finding correctness bugs},
  author       = {Kim, Seulbae and Kim, Taesoo},
  year         = 2022,
  booktitle    = esec
}

@inproceedings{lee2022towards,
  title        = {Towards Safe, Realistic Testbed for Robotic Systems with Human Interaction},
  author       = {Lee, Bhoram and Brookshire, Jonathan and Yahata, Rhys and Samarasekera, Supun},
  year         = 2022,
  booktitle    = icra
}

@Article{liu2021ocrtoc,
  author  = {Liu, Ziyuan and Liu, Wei and Qin, Yuzhe and Xiang, Fanbo and Gou, Minghao and Xin, Songyan and Roa, Maximo A. and Calli, Berk and Su, Hao and Sun, Yu and Tan, Ping},
  journal = ral,
  title   = {OCRTOC: A Cloud-Based Competition and Benchmark for Robotic Grasping and Manipulation},
  year    = {2022},
  number  = {1},
  volume  = {7},
}

@article{lundberg2022robotics,
  title        = {Robotic Nursing Assistant Applications and Human Subject Tests through Patient Sitter and Patient Walker Tasks},
  author       = {Lundberg, Cody Lee and Sevil, Hakki Erhan and Behan, Deborah and Popa, Dan O.},
  year         = 2022,
  journal      = robotics,
  VOLUME = {11},
  YEAR = {2022},
  NUMBER = {3},
}

@article{marrella2022methodology,
  title        = {A Methodology to Design and Evaluate HRI Teaming Tasks in Robotic Competitions},
  author       = {Marrella, Andrea and Wang, Lun and Iocchi, Luca and Nardi, Daniele},
  year         = 2022,
  journal      = jhri,
  volume = {11},
  number = {3},
  numpages = {22},
}

@inproceedings{mayoral2022iros,
  title        = {SROS2: Usable Cyber Security Tools for ROS 2},
  author       = {Mayoral-Vilches, Victor and White, Ruffin and Caiazza, Gianluca and Arguedas, Mikael},
  year         = 2022,
  booktitle    = iros
}

@inproceedings{okwu2022development,
  title        = {Development of a Light Weight Autonomous Lawn Mower and Performance Analysis using Fuzzy Logic Technique},
  author       = {Okwu, Modestus O. and Tartibu, Lagouge K. and Enarevba, Dolor Roy and Oyejide, Oluwayomi J. and Otanocha, Omonigho B. and Adumene, Sidum},
  year         = 2022,
  booktitle    = icabcd
}

@inproceedings{ortega2022iros,
  title        = {Testing Service Robots in the Field: An Experience Report},
  author       = {Ortega, Argentina and Hochgeschwender, Nico and Berger, Thorsten},
  year         = 2022,
  booktitle    = iros
}

@inproceedings{paez2022pedestrian,
  title        = {Pedestrian-robot interactions on autonomous crowd navigation: Reactive control methods and evaluation metrics},
  author       = {Paez-Granados, Diego and He, Yujie and Gonon, David and Jia, Dan and Leibe, Bastian and Suzuki, Kenji and Billard, Aude},
  year         = 2022,
  booktitle    = iros
}

@Article{pak2022field,
  author  = {Pak, Jeonghyeon and Kim, Jeongeun and Park, Yonghyun and Son, Hyoung Il},
  journal = access,
  title   = {Field Evaluation of Path-Planning Algorithms for Autonomous Mobile Robot in Smart Farms},
  year    = {2022},
  volume  = {10},
}

@inproceedings{palermino2022development,
  title        = {Development and testing of a collision avoidance algorithm for industrial applications},
  author       = {Palermino, Lorenzo and Fathy, Ahmed Mohsen Mohamed and Carnevale, Marco and Giberti, Hermes},
  year         = 2022,
  booktitle    = mesa
}

@inproceedings{pang2022mdpfuzz,
  title        = {MDPFuzz: testing models solving Markov decision processes},
  author       = {Pang, Qi and Yuan, Yuanyuan and Wang, Shuai},
  year         = 2022,
  booktitle    = issta
}

@Article{platt2022comparative,
  author  = {Platt, Jonathan and Ricks, Kenneth},
  journal = jirs,
  title   = {Comparative analysis of ros-unity3d and ros-gazebo for mobile ground robot simulation},
  year    = {2022},
  number  = {4},
  volume  = {106},
}

@inproceedings{sadek2022depth,
  title        = {An in-depth experimental study of sensor usage and visual reasoning of robots navigating in real environments},
  author       = {Sadek, Assem and Bono, Guillaume and Chidlovskii, Boris and Wolf, Christian},
  year         = 2022,
  booktitle    = icra
}

@inproceedings{santos2022schema,
  title        = {Schema-guided testing of message-oriented systems},
  author       = {Santos, Andr{\'e} and Cunha, Alcino and Macedo, Nuno},
  year         = 2022,
  journal      = enase
}

@InProceedings{santos2022irc,
  author    = {Santos, Marcela G. Dos and Petrillo, Fabio and Hall{\'{e}}, Sylvain and Gu{\'{e}}h{\'{e}}neuc, Yann-Ga{\"{e}}l},
  booktitle = irc,
  title     = {An approach to apply Automated Acceptance Testing for Industrial Robotic Systems},
  year      = {2022},
}

@inproceedings{sartori2022integration,
  title        = {Integration of test generation into simulation-based platforms: an experience report},
  author       = {Sartori, Luca Vittorio and Guiochet, J\'{e}r\'{e}mie and Waeselynck, H\'{e}l\`{e}ne and Galvan, Aizar Antonio Berlanga and H\'{e}bert-Vernhes, Simon and Albert, Magnus},
  year         = 2022,
  booktitle    = ast
}

@inproceedings{shin2022hands,
  title        = {Hands-On Field Operational Test Dataset of a Multi-Controller CPS: A Modeled Case Study on Autonomous Driving},
  author       = {Shin, Yong-Jun and Cho, Esther and Kim, Hansu and Bae, Doo-Hwan},
  year         = 2022,
  booktitle    = sose
}

@Article{sobczak2022building,
  author  = {Sobczak, {{\L}}ukasz and Filus, Katarzyna and Doma{{\'{n}}}ska, Joanna and Doma{{\'{n}}}ski, Adam},
  journal = sensors,
  title   = {Building a Real-Time Testing Platform for Unmanned Ground Vehicles with UDP Bridge},
  year    = {2022},
  number  = {21},
  volume  = {22},
}

@article{touzout2022mobile,
  title        = {Mobile robot energy modelling integrated into ros and gazebo-based simulation environment},
  author       = {Touzout, Walid and Benazzouz, Djamel and Benmoussa, Yahia},
  year         = 2022,
  journal      = msc,
  volume={50},
  number={2},
  pages={}
}

@Article{wang2022toward,
  author  = {Wang, Bingqing and Wang, Rui and Song, Houbing},
  journal = tii,
  title   = {Toward the trustworthiness of industrial robotics using differential fuzz testing},
  year    = {2022},
  number  = {3},
  volume  = {19},
}

@inproceedings{watanabe2022hardware,
  title        = {Hardware Design and Tests of Two-Wheeled Robot Platform for Searching Survivors in Debris Cones},
  author       = {Watanabe, Masahiro and Ozawa, Yu and Takahashi, Kenichi and Kimura, Tetsuya and Tadakuma, Kenjiro and Marafioti, Giancarlo and Tadokoro, Satoshi},
  year         = 2022,
  booktitle    = ssrr
}

@inproceedings{weng2022safety,
  title        = {On safety testing, validation, and characterization with scenario-sampling: A case study of legged robots},
  author       = {Weng, Bowen and Castillo, Guillermo A and Zhang, Wei and Hereid, Ayonga},
  year         = 2022,
  booktitle    = iros
}

@inproceedings{xie20222icra,
  title        = {{{ROZZ}}: {{Property-based Fuzzing}} for {{Robotic Programs}} in {{ROS}}},
  author       = {Xie, Kai-Tao and Bai, Jia-Ju and Zou, Yong-Hao and Wang, Yu-Ping},
  year         = 2022,
  booktitle    = icra
}

@Article{zhou2022design,
  author  = {Zhou, Hang and Zhang, Shunlu and Zhang, Junxiong and Zhang, Chunlong and Wang, Song and Zhai, Yihao and Li, Wei},
  journal = jfr,
  title   = {Design, development, and field evaluation of a rubber tapping robot},
  year    = {2022},
  number  = {1},
  volume  = {39},
}

@TechReport{iso29119pt4,
  author = {{IEEE/ISO/IEC }},
  title  = {{Software and systems engineering--Software testing -- Part 4: Test techniques}},
  year   = {2021},
  number = {ISO/IEC/IEEE 29119-4},
  type   = {Int. Std.},
}

@inproceedings{afzal2021icst,
  title        = {Simulation for Robotics Test Automation: Developer Perspectives},
  author       = {Afzal, Afsoon and Katz, Deborah S. and Le Goues, Claire and Timperley, Christopher Steven},
  year         = 2021,
  booktitle    = icst,
}

@Article{hietanen2021benchmarking,
  author  = {Antti Hietanen and Jyrki Latokartano and Alessandro Foi and Roel Pieters and Ville Kyrki and Minna Lanz and Joni-Kristian K{\"{a}}m{\"{a}}r{\"{a}}inen},
  journal = ras,
  title   = {Benchmarking pose estimation for robot manipulation},
  year    = {2021},
  volume  = {143},
}

@Article{bertolino2021survey,
  author  = {Bertolino, Antonia and Braione, Pietro and Angelis, Guglielmo De and Gazzola, Luca and Kifetew, Fitsum and Mariani, Leonardo and Orr\`{u}, Matteo and Pezz\`{e}, Mauro and Pietrantuono, Roberto and Russo, Stefano and Tonella, Paolo},
  journal = {CSUR},
  title   = {A Survey of Field-based Testing Techniques},
  year    = {2021},
  number  = {5},
  volume  = {54},
}

@Article{WANG2021101838,
  author  = {Chundong Wang and Yee Ching Tok and Rohini Poolat and Sudipta Chattopadhyay and Mohan Rajesh Elara},
  journal = jsa,
  title   = {How to secure autonomous mobile robots? An approach with fuzzing, detection and mitigation},
  year    = {2021},
  volume  = {112},
}

@article{chung2021hardware,
  AUTHOR = {Chung, Yi and Yang, Yee-Pien},
  TITLE = {Hardware-in-the-Loop Simulation of Self-Driving Electric Vehicles by Dynamic Path Planning and Model Predictive Control},
  JOURNAL = {Electronics},
  VOLUME = {10},
  YEAR = {2021},
  NUMBER = {19},
}

@inproceedings{delgado2021fuzz,
  title        = {Fuzz testing in behavior-based robotics},
  author       = {Delgado, Rodrigo and Campusano, Miguel and Bergel, Alexandre},
  year         = 2021,
  booktitle    = icra
}

@Article{santos2021arxiv,
  author  = {dos Santos, Marcela G. and Petrillo, Fabio},
  journal = arxiv,
  title   = {Software Engineering for Robotic Systems: A Systematic Mapping Study},
  year    = {2021},
  number  = {2102.12520},
}

@article{douthwaite2021modular,
  title        = {A modular digital twinning framework for safety assurance of collaborative robotics},
  author       = {Douthwaite, James A and Lesage, Benjamin and Gleirscher, Mario and Calinescu, Radu and Aitken, Jonathan M and Alexander, Rob and Law, James},
  year         = 2021,
  VOLUME       = {8},
  journal      = frontiers-ai
}

@inproceedings{harbin2021model,
  title        = {Model-driven simulation-based analysis for multi-robot systems},
  author       = {Harbin, James and Gerasimou, Simos and Matragkas, Nicholas and Zolotas, Athanasios and Calinescu, Radu},
  year         = 2021,
  booktitle    = models
}

@InProceedings{hereau2021fault,
  author    = {Hereau, Adrien and Godary-Dejean, Karen and Guiochet, J{\'{e}}r{\'{e}}mie and Crestani, Didier},
  booktitle = icra,
  title     = {A Fault Tolerant Control Architecture Based on Fault Trees for an Underwater Robot Executing Transect Missions},
  year      = {2021},
}

@inproceedings{kaneshige2021overcome,
  title        = {How to Overcome the Difficulties in Programming and Debugging Mobile Social Robots?},
  author       = {Kaneshige, Yuya and Satake, Satoru and Kanda, Takayuki and Imai, Michita},
  year         = 2021,
  booktitle    = hri
}

@article{lestingi2021deployment,
  title        = {A Deployment Framework for Formally Verified Human-Robot Interactions},
  author       = {Lestingi, Livia and Askarpour, Mehrnoosh and Bersani, Marcello M. and Rossi, Matteo},
  year         = 2021,
  journal      = access,
  volume={9},
  number={},
}

@inproceedings{luperto2021my,
  title        = {What is my Robot Doing? Remote Supervision to Support Robots for Older Adults Independent Living: a Field Study},
  author       = {Luperto, Matteo and Romeo, Marta and Monroy, Javier and Vuono, Alessandro and Basilico, Nicola and Gonzalez-Jimenez, Javier and Borghese, N. Alberto},
  year         = 2021,
  booktitle    = ecmr
}

@Article{gadaleta2021extensive,
  author   = {M. Gadaleta and G. Berselli and M. Pellicciari and F. Grassia},
  journal  = {Rob. Comput. Integr. Manuf.},
  title    = {Extensive experimental investigation for the optimization of the energy consumption of a high payload industrial robot with open research dataset},
  year     = {2021},
  volume   = {68},
  fjournal = {Robotics and Computer-Integrated Manufacturing},
}

@inproceedings{moshayedi2021simulationPID,
  title        = {Simulation study and PID Tune of Automated Guided Vehicles (AGV)},
  author       = {Moshayedi, Ata Jahangir and Li, Jinsong and Liao, Liefa},
  year         = 2021,
  booktitle    = civemsa
}

@article{nagrath2021smartts,
  title        = {SmartTS: A Component-Based and Model-Driven Approach to Software Testing in Robotic Software Ecosystem},
  author       = {Nagrath, Vineet and Schlegel, Christian},
  year         = 2021,
  journal      = ijcasa,
  volume={12},
  number={7},
}

@inproceedings{reithner2021analysis,
  title        = {Analysis Of The Interaction Between Safety And Security Demonstrated On A Mobile Robot And A Production Network.},
  author       = {Reithner, Isabella and Papa, Maximilian and Aburaia, Mohamed and W{\"o}ber, Wilfried and Ambros, Clemens},
  year         = 2021,
  booktitle    = daaam
}

@inproceedings{santos2021rose,
  title        = {The {{High-Assurance ROS Framework}}},
  author       = {Santos, Andr{\'e} and Cunha, Alcino and Macedo, Nuno},
  year         = 2021,
  booktitle    = rose
}

@inproceedings{sohail2021ecmr,
  title        = {Property-{{Based Testing}} in {{Simulation}} for {{Verifying Robot Action Execution}} in {{Tabletop Manipulation}}},
  author       = {Sohail, Salman Omar and Mitrevski, Alex and Hochgeschwender, Nico and Ploger, Paul G},
  year         = 2021,
  booktitle    = ecmr
}

@inproceedings{song2021wain,
  title        = {Concepts in {{Testing}} of {{Autonomous Systems}}: {{Academic Literature}} and {{Industry Practice}}},
  author       = {Song, Qunying and Engstr{\"o}m, Emelie and Runeson, Per},
  year         = 2021,
  booktitle    = wain
}

@InProceedings{suarez2021experimental,
  author    = {Suarez, Alejandro and Romero, Honorio and Salmoral, Rafael and Acosta, Jos{\'{e}} Alberto and Zambrano, Jes{\'{u}}s and Ollero, Anibal},
  booktitle = airpharo,
  title     = {Experimental Evaluation of Aerial Manipulation Robot for the Installation of Clip Type Bird Diverters: Outdoor Flight Tests},
  year      = {2021},
}

@inproceedings{swanborn2021icse,
  title        = {Robot {{Runner}}: {{A Tool}} for {{Automatically Executing Experiments}} on {{Robotics Software}}},
  author       = {Swanborn, Stan and Malavolta, Ivano},
  year         = 2021,
  booktitle    = icse-companion
}

@InProceedings{teixeira2021cloud,
  author    = {Teixeira, S{\'{e}}rgio and Arrais, Rafael and Veiga, Germano},
  booktitle = indin,
  title     = {Cloud Simulation for Continuous Integration and Deployment in Robotics},
  year      = {2021},
}

@inproceedings{woodlief2021fuzzing,
  title        = {Fuzzing Mobile Robot Environments for Fast Automated Crash Detection},
  author       = {Woodlief, Trey and Elbaum, Sebastian and Sullivan, Kevin},
  year         = 2021,
  booktitle    = icra
}

@Article{xu2021four,
  author  = {Xu, Tong and Wang, Dong and Xiao, Zuodong and Chu, Cancan and Zhang, Weigong},
  journal = ijars,
  title   = {A four-level test system for evaluating pavement compaction performance of autonomous articulated vehicles},
  year    = {2021},
  number  = {1},
  volume  = {18},
}

@Article{Deng01012021,
  author  = {Xuefeng Deng and Bingqian Zhou and Xinyi Sun and Hua Yang and Lingyu Chen},
  journal = {Systems Science \& Control Engineering},
  title   = {A method for reliability detection of automated guided vehicle based on timed automata},
  year    = {2021},
  number  = {1},
  volume  = {9},
}

@inproceedings{afzal2020icst,
  title        = {A Study on Challenges of Testing Robotic Systems},
  author       = {Afzal, Afsoon and Goues, Claire Le and Hilton, Michael and Timperley, Christopher Steven},
  year         = 2020,
  booktitle    = icst,
}

@inproceedings{aliabadi2020ikt,
  title        = {Challenges of Specification Mining-Based Test Oracle for Cyber-Physical Systems},
  author       = {Aliabadi, Maryam Raiyat and Haghighi, Hassan and Asl, Mojtaba Vahidi and Meybodi, Ramak Ghavamizadeh},
  year         = 2020,
  booktitle    = ikt,
}

@Article{arango2020drive,
  author   = {Arango, J. Felipe and Bergasa, Luis M. and Revenga, Pedro A. and Barea, Rafael and L{\'{o}}pez-Guill{\'{e}}n, Elena and G{\'{o}}mez-Hu{\'{e}}lamo, Carlos and Araluce, Javier and Guti{\'{e}}rrez, Rodrigo},
  journal  = {Sensors-basel.},
  title    = {Drive-By-Wire Development Process Based on ROS for an Autonomous Electric Vehicle},
  year     = {2020},
  number   = {21},
  volume   = {20},
  fjournal = {Sensors},
}

@Article{babic2020vehicle,
  author  = {Babi{\'{c}}, Anja and Vasiljevi{\'{c}}, Goran and Mi{\v{s}}kovi{\'{c}}, Nikola},
  journal = {RA-L},
  title   = {Vehicle-in-the-Loop Framework for Testing Long-Term Autonomy in a Heterogeneous Marine Robot Swarm},
  year    = {2020},
  number  = {3},
  volume  = {5},
}

@inproceedings{bashetty2020deepcrashtest,
  title        = {DeepCrashTest: Turning Dashcam Videos into Virtual Crash Tests for Automated Driving Systems},
  author       = {Bashetty, Sai Krishna and Ben Amor, Heni and Fainekos, Georgios},
  year         = 2020,
  booktitle    = icra
}

@Article{bi2020rcim,
  author   = {Z.M. Bi and Zhonghua Miao and Bin Zhang and Chris W.J. Zhang},
  journal  = {Rob. Comput. Integr. Manuf.},
  title    = {The state of the art of testing standards for integrated robotic systems},
  year     = {2020},
  volume   = {63},
  fjournal = {Robotics and Computer-Integrated Manufacturing},
}

@Article{birrell2020field,
  author  = {Birrell, S. and Hughes, J. and Cai, J. Y. and Iida, F.},
  journal = jfr,
  title   = {A field-tested robotic harvesting system for iceberg lettuce},
  year    = {2020},
  number  = {2},
  volume  = {37},
}

@inproceedings{breitenhuber2020towards,
  title        = {Towards application level testing of ROS networks},
  author       = {Breitenhuber, Guido},
  year         = 2020,
  booktitle    = irc
}

@Article{brito2020springer,
  author  = {Brito, Maria A. S. and Souza, Simone R. S. and Souza, Paulo S. L.},
  journal = sqj,
  title   = {Integration Testing for Robotic Systems},
  year    = {2022},
  number  = {1},
  volume  = {30},
}

@Article{cardoso2020towards,
  author  = {Cardoso, Rafael C. and Dennis, Louise A. and Farrell, Marie and Fisher, Michael and Luckcuck, Matt},
  journal = {EPTCS},
  title   = {Towards Compositional Verification for Modular Robotic Systems},
  year    = {2020},
  volume  = {329},
}

@inproceedings{casiddu2020humanoid,
  title        = {Humanoid Robotics: Guidelines for Usability},
  author       = {Casiddu, Niccol{\`o} and Burlando, Francesco and Vacanti, Annapaola},
  year         = 2020,
  booktitle    = ihsed
}

@Article{falco2020benchmarking,
  author  = {Falco, Joe and Hemphill, Daniel and Kimble, Kenny and Messina, Elena and Norton, Adam and Ropelato, Rafael and Yanco, Holly},
  journal = ral,
  title   = {Benchmarking Protocols for Evaluating Grasp Strength, Grasp Cycle Time, Finger Strength, and Finger Repeatability of Robot End-Effectors},
  year    = {2020},
  number  = {2},
  volume  = {5},
}

@inproceedings{garcia2020esec,
  title        = {Robotics Software Engineering: A Perspective from the Service Robotics Domain},
  author       = {Garc{\'i}a, Sergio and Str{\"u}ber, Daniel and Brugali, Davide and Berger, Thorsten and Pelliccione, Patrizio},
  year         = {2020},
  booktitle    = {FSE}
}

@incollection{gotlieb2020testing,
  title        = {Testing industrial robotic systems: A new battlefield!},
  author       = {Gotlieb, Arnaud and Marijan, Dusica and Spieker, Helge},
  year         = 2020,
  booktitle    = ser
}

@article{guo2020precision,
  title        = {Precision Landing Test and Simulation of the Agricultural UAV on Apron},
  author       = {Guo, Yangyang and Guo, Jiaqian and Liu, Chang and Xiong, Hongting and Chai, Lilong and He, Dongjian},
  year         = 2020,
  journal      = sensors,
  VOLUME = {20},
  YEAR = {2020},
  NUMBER = {12},

}

@Article{seyyedhasani2020collaboration,
  author  = {Hasan Seyyedhasani and Chen Peng and Wei-jiunn Jang and Stavros G. Vougioukas},
  journal = compag,
  title   = {Collaboration of human pickers and crop-transporting robots during harvesting – Part I: Model and simulator development},
  year    = {2020},
  volume  = {172},
}

@inproceedings{huck2020simulation,
  title        = {Simulation-based testing for early safety-validation of robot systems},
  author       = {Huck, Tom P and Ledermann, Christoph and Kr{\"o}ger, Torsten},
  year         = 2020,
  booktitle    = spce
}

@inproceedings{insam2020high,
  title        = {High Fidelity Real-Time Hybrid Substructure Testing Using Iterative Learning Control},
  author       = {Insam, Christina and Kist, Arian and Rixen, Daniel J.},
  year         = 2020,
  booktitle    = isr
}

@Article{kanter2020rie,
  author  = {Kanter, Gert and Vain, J{\"u}ri},
  journal = jrie,
  title   = {Model-Based Testing of Autonomous Robots Using TestIt},
  year    = {2020},
  number  = {1},
  volume  = {6},
}

@inproceedings{katz20202icra,
  title        = {Detecting {{Execution Anomalies As}} an {{Oracle}} for {{Autonomy Software Robustness}}},
  author       = {Katz, Deborah S. and Hutchison, Casidhe and Zizyte, Milda and Goues, Claire Le},
  year         = 2020,
  booktitle    = icra
}

@Article{kimble2020benchmarking,
  author  = {Kimble, K. and Van Wyk, K. and Falco, J. and Messina, E. and Sun, Y. and Shibata, M. and Uemura, W. and Yokokohji, Y.},
  journal = ral,
  title   = {Benchmarking Protocols for Evaluating Small Parts Robotic Assembly Systems},
  year    = {2020},
  number  = {2},
  volume  = {5},
}

@article{luckcuck2020csur,
  title        = {Formal Specification and Verification of Autonomous Robotic Systems: A Survey},
  author       = {Luckcuck, Matt and Farrell, Marie and Dennis, Louise and Dixon, Clare and Fisher, Michael},
  year         = 2020,
  journal      = acmcs,
  volume = {52},
  number = {5},
  numpages = {41},
}

@InProceedings{maftei2020multi,
  author    = {Maftei, {{\c{S}}}tefan and Novischi, Dan},
  booktitle = aqtr,
  title     = {Multi-Robot 2D Simulator for the evaluation of self-organizing robot swarms},
  year      = {2020},
}

@Article{mannion2020introducing,
  author  = {Mannion, Arlene and Summerville, Sarah and Barrett, Eva and Burke, Megan and Santorelli, Adam and Kruschke, Cheryl and Felzmann, Heike and Kovacic, Tanja and Murphy, Kathy and Casey, Dympna and others},
  journal = ijsr,
  title   = {Introducing the social robot MARIO to people living with dementia in long term residential care: reflections},
  year    = {2020},
  number  = {2},
  volume  = {12},
}

@Article{webster2020corroborative,
  author  = {Matt Webster and David Western and Dejanira Araiza-Illan and Clare Dixon and Kerstin Eder and Michael Fisher and Anthony G Pipe},
  journal = ijrr,
  title   = {A corroborative approach to verification and validation of human–robot teams},
  year    = {2020},
  number  = {1},
  volume  = {39},
}

@inproceedings{megalingam2020comparison,
  title        = {Comparison of Planned Path and Travelled Path Using ROS Navigation Stack},
  author       = {Megalingam, Rajesh Kannan and Rajendraprasad, Anandu and Manoharan, Sakthiprasad Kuttankulangara},
  year         = 2020,
  booktitle    = incet
}

@inproceedings{mercier2020terrestrial,
  title        = {Terrestrial Testing of Multi-Agent, Relative Guidance, Navigation, and Control Algorithms},
  author       = {Mercier, Mark and Phillips, Sean and Shubert, Matt and Dong, Wenjie},
  year         = 2020,
  booktitle    = plans
}

@Article{Tanaka17012020,
  author  = {Motoyasu Tanaka and Kazuyuki Kon and Mizuki Nakajima and Nobutaka Matsumoto and Shinnosuke Fukumura and Kosuke Fukui and Hidemasa Sawabe and Masahiro Fujita and Kenjiro Tadakuma},
  journal = adv-robotics,
  title   = {Development and field test of the articulated mobile robot T2 Snake-4 for plant disaster prevention},
  year    = {2020},
  number  = {2},
  volume  = {34},
}

@inproceedings{pinrath2020integration,
  title        = {Integration of Virtual Robot Environmental Platform with Robot Operating System for Real-time Simulation of A Robot},
  author       = {Pinrath, Nattawat and Matsuhira, Nobuto},
  year         = 2020,
  booktitle    = sice
}

@inproceedings{robert2020irc,
  title        = {Testing a Non-Deterministic Robot in Simulation - {{How}} Many Repeated Runs?},
  author       = {Robert, Clement and Guiochet, Jeremie and Waeselynck, Helene},
  year         = 2020,
  booktitle    = irc
}

@Article{robert2020ese,
  author  = {Robert, Cl{\'e}ment and Sotiropoulos, Thierry and Waeselynck, H{\'e}l{\`e}ne and Guiochet, J{\'e}r{\'e}mie and Vernhes, Simon},
  journal = ese,
  title   = {The Virtual Lands of {{Oz}}: Testing an Agribot in Simulation},
  year    = {2020},
  number  = {3},
  volume  = {25},
}

@inproceedings{sadka2020virtual,
  title        = {Virtual-reality as a Simulation Tool for Non-humanoid Social Robots},
  author       = {Sadka, Ofir and Giron, Jonathan and Friedman, Doron and Zuckerman, Oren and Erel, Hadas},
  year         = 2020,
  booktitle    = ea-chi
}

@Article{seyyedhasani2020collaborationPartII,
  author  = {Seyyedhasani, Hasan and Peng, Chen and Jang, Wei-jiunn and Vougioukas, Stavros G},
  journal = compag,
  title   = {Collaboration of human pickers and crop-transporting robots during harvesting--Part II: Simulator evaluation and robot-scheduling case-study},
  year    = {2020},
  volume  = {172},
}

@InProceedings{shell2020reality,
  author    = {Shell, Dylan A. and O{\textquoteright}Kane, Jason M.},
  booktitle = icra,
  title     = {Reality as a simulation of reality: robot illusions, fundamental limits, and a physical demonstration},
  year      = {2020},
}

@InProceedings{verdu2020experimental,
  author    = {Verdu, Titouan and Maury, Nicolas and Narvor, Pierre and Seguin, Florian and Roberts, Gregory and Couvreux, Fleur and Cayez, Gr{\'{e}}goire and Bronz, Murat and Hattenberger, Gautier and Lacroix, Simon},
  booktitle = iros,
  title     = {Experimental flights of adaptive patterns for cloud exploration with UAVs},
  year      = {2020},
}

@phdthesis{wang2020reliability,
  title        = {Reliability Engineering for Long-term Deployment of Autonomous Service Robots},
  author       = {Wang, Shengye},
  year         = 2020,
  publisher    = {University of California San Diego}
}

@inproceedings{wolf2020evolution,
  title        = {Evolution of Robotic Simulators: Using UE 4 to Enable Real-World Quality Testing of Complex Autonomous Robots in Unstructured Environments},
  author       = {Wolf, Patrick and Groll, Tobias and Hemer, Steffen and Berns, Karsten},
  year         = 2020,
  booktitle    = simultech
}

@Article{xiong2020autonomous,
  author  = {Xiong, Ya and Ge, Yuanyue and Grimstad, Lars and From, P{{\aa}}l J.},
  journal = jfr,
  title   = {An autonomous strawberry-harvesting robot: Design, development, integration, and field evaluation},
  year    = {2020},
  number  = {2},
  volume  = {37},
}

@InProceedings{zofka2020pushing,
  author    = {Zofka, Marc Ren{\'{e}} and T{\"{o}}ttel, Lars and Zipfl, Maximilian and Heinrich, Marc and Fleck, Tobias and Schulz, Patrick and Z{\"{o}}llner, J. Marius},
  booktitle = mfi,
  title     = {Pushing ROS towards the Dark Side: A ROS-based Co-Simulation Architecture for Mixed-Reality Test Systems for Autonomous Vehicles},
  year      = {2020},
}

@Article{banerjee2018integrated,
  author  = {Banerjee, Debdeep and Yu, Kevin},
  journal = {IEEE Access},
  title   = {Integrated Test Automation for Evaluating a Motion-Based Image Capture System Using a Robotic Arm},
  year    = {2019},
  volume  = {7},
}

@Article{bozhinoski2019jjss,
  author  = {Darko Bozhinoski and Davide {Di Ruscio} and Ivano Malavolta and Patrizio Pelliccione and Ivica Crnkovic},
  journal = {JSS},
  title   = {Safety for mobile robotic systems: A systematic mapping study from a software engineering perspective},
  year    = {2019},
  volume  = {151},
}

@inproceedings{collet2019stress,
  title        = {Stress Testing of Single-Arm Robots Through Constraint-Based Generation of Continuous Trajectories},
  author       = {Collet, Mathieu and Gotlieb, Arnaud and Lazaar, Nadjib and Mossige, Morten},
  year         = 2019,
  booktitle    = aitest
}

@Article{eckert2019benchmarking,
  author  = {Eckert, Peter and Ijspeert, Auke J.},
  journal = tro,
  title   = {Benchmarking Agility For Multilegged Terrestrial Robots},
  year    = {2019},
  number  = {2},
  volume  = {35},
}

@inproceedings{erich2019design,
  title        = {Design and development of a physical integration testing framework for robotic manipulators},
  author       = {Erich, Floris and Saksena, Abhilasha and Biggs, Geoffrey and Ando, Noriaki},
  year         = 2019,
  booktitle    = sii
}

@inproceedings{fernandez2019securing,
  title        = {Securing UAV communications using ROS with custom ECIES-based method},
  author       = {Fernandez, Manuel J. and Sanchez-Cuevas, Pedro J. and Heredia, Guillermo and Ollero, Anibal},
  year         = 2019,
  booktitle    = red-uas
}

@inproceedings{kanter2019dessert,
  title        = {{{TestIt}}: An {{Open-Source Scalable Long-Term Autonomy Testing Toolkit}} for {{ROS}}},
  author       = {Kanter, Gert and Vain, Juri},
  year         = 2019,
  booktitle    = dessert
}

@inproceedings{kim2019rvfuzzer,
  title        = {RVFUZZER: Finding Input Validation Bugs in Robotic Vehicles Through Control-Guided Testing},
  author       = {Kim, Taegyu and Kim, Chung Hwan and Rhee, Junghwan and Fei, Fan and Tu, Zhan and Walkup, Gregory and Zhang, Xiangyu and Deng, Xinyan and Xu, Dongyan},
  year         = 2019,
  booktitle    = usenix
}

@inproceedings{li2019uav,
  title        = {UAV System Integration of Real-time Sensing and Flight Task Control for Autonomous Building Inspection Task},
  author       = {Li, Gong-Yi and Soong, Ru-Tai and Liu, Jyi-Shane and Huang, Yen-Ting},
  year         = 2019,
  booktitle    = taai
}

@Article{liaqat2019autonomous,
  author  = {Liaqat, A. and Hutabarat, W. and Tiwari, D. and Tinkler, L. and Harra, D. and Morgan, B. and Taylor, A. and Lu, T. and Tiwari, A.},
  journal = ijamt,
  title   = {Autonomous mobile robots in manufacturing: Highway Code development, simulation, and testing},
  year    = {2019},
  number  = {9},
  volume  = {104},
}

@inproceedings{munawar2019real,
  title        = {A Real-Time Dynamic Simulator and an Associated Front-End Representation Format for Simulating Complex Robots and Environments},
  author       = {Munawar, Adnan and Wang, Yan and Gondokaryono, Radian and Fischer, Gregory S.},
  year         = 2019,
  booktitle    = iros
}

@article{norris2019system,
  title        = {System-level testing and evaluation plan for field robots: A tutorial with test course layouts},
  author       = {Norris, William R and Patterson, Albert E},
  year         = 2019,
  journal      = robotics,
  VOLUME = {8},
  NUMBER = {4},
  ARTICLE-NUMBER = {83},
}

@Article{norton2019standard,
  author  = {Norton, Adam and Gavriel, Peter and Yanco, Holly},
  journal = ssms,
  title   = {A Standard Test Method for Evaluating Navigation and Obstacle Avoidance Capabilities of AGVs and AMRs},
  year    = {2019},
  number  = {2},
  volume  = {3},
}

@Article{punzo2019bipartite,
  author  = {Punzo, Giuliano and MacLeod, Charles and Baumanis, Kristaps and Summan, Rahul and Dobie, Gordon and Pierce, Gareth and Macdonald, Malcolm},
  journal = jirs,
  title   = {Bipartite guidance, navigation and control architecture for autonomous aerial inspections under safety constraints},
  year    = {2019},
  number  = {3},
  volume  = {95},
}

@inproceedings{queiroz2019geoscenario,
  title        = {GeoScenario: An open DSL for autonomous driving scenario representation},
  author       = {Queiroz, Rodrigo and Berger, Thorsten and Czarnecki, Krzysztof},
  year         = 2019,
  booktitle    = iv
}

@inproceedings{sarkar2019behavior,
  title        = {A behavior driven approach for sampling rare event situations for autonomous vehicles},
  author       = {Sarkar, Atrisha and Czamecki, Krzysztof},
  year         = 2019,
  booktitle    = iros
}

@inproceedings{satoh2019developing,
  title        = {Developing and Testing Networked Software for Moving Robots.},
  author       = {Satoh, Ichiro},
  year         = 2019,
  booktitle    = enase
}

@inproceedings{schiegg2019novel,
  title        = {A Novel Simulation Framework for the Design and Testing of Advanced Driver Assistance Systems},
  author       = {Schiegg, Florian A. and Krost, Johannes and Jesenski, Stefan and Frye, Johannes},
  year         = 2019,
  booktitle    = vtc
}

@Article{selecky2019analysis,
  author  = {Seleck{\`y}, Martin and Faigl, Jan and Rollo, Milan},
  journal = jirs,
  title   = {Analysis of using mixed reality simulations for incremental development of multi-uav systems},
  year    = {2019},
  number  = {1},
  volume  = {95},
}

@inproceedings{vieira2019copadrive,
  title        = {COPADRIVe - A Realistic Simulation Framework for Cooperative Autonomous Driving Applications},
  author       = {Vieira, Bruno and Severino, Ricardo and Filho, Enio Vasconcelos and Koubaa, Anis and Tovar, Eduardo},
  year         = 2019,
  booktitle    = iccve
}

@Article{krueger2019testing,
  author  = {Volker Krueger and Francesco Rovida and Bjarne Grossmann and Ronald Petrick and Matthew Crosby and Arnaud Charzoule and German {Martin Garcia} and Sven Behnke and Cesar Toscano and Germano Veiga},
  journal = rcim,
  title   = {Testing the vertical and cyber-physical integration of cognitive robots in manufacturing},
  year    = {2019},
  volume  = {57},
}

@article{cwikakala2019testing,
  title        = {Testing procedure of unmanned aerial vehicles (UAVs) trajectory in automatic missions},
  author       = {{\'C}wi{\k{a}}ka{\l}a, Pawe{\l}},
  JOURNAL = {Applied Sciences},
  VOLUME = {9},
  YEAR = {2019},
  NUMBER = {17},
}

@Article{xiong2019development,
  author  = {Ya Xiong and Cheng Peng and Lars Grimstad and P{{\aa}}l Johan From and Volkan Isler},
  journal = compag,
  title   = {Development and field evaluation of a strawberry harvesting robot with a cable-driven gripper},
  year    = {2019},
  volume  = {157},
}

@inproceedings{yoon2019analysis,
  title        = {Analysis of Automatic through Autonomous - Unmanned Ground Vehicles (A-UGVs) Towards Performance Standards},
  author       = {Yoon, Soocheol and Bostelman, Roger},
  year         = 2019,
  booktitle    = rose
}

@inproceedings{zhang2019testing,
  title        = {Testing Graph Searching Based Path Planning Algorithms by Metamorphic Testing},
  author       = {Zhang, Jiantao and Zheng, Zheng and Yin, Beibei and Qiu, Kun and Liu, Yang},
  year         = 2019,
  booktitle    = prdc
}

@Misc{istqb2018,
  author = {{International Software Testing Qualifications Board (ITSQB)}},
  title  = {{Certified Tester -- Foundation Level Syllabus}},
  year   = {2024},
}

@Article{chowdhury2018experiments,
  author   = {Abhra Roy Chowdhury and G.S. Soh and S.H. Foong and K.L. Wood},
  journal  = {Robot. Auton. Syst.},
  title    = {Experiments in robust path following control of a rolling and spinning robot on outdoor surfaces},
  year     = {2018},
  volume   = {106},
  fjournal = {Robotics and Autonomous Systems},
}

@Article{conte2018development,
  author   = {Conte, Giuseppe and Scaradozzi, David and Mannocchi, Daniele and Raspa, Paolo and Panebianco, Luca and Screpanti, Laura},
  journal  = {JINT},
  title    = {Development and experimental tests of a ROS multi-agent structure for autonomous surface vehicles},
  year     = {2018},
  number   = {3},
  volume   = {92},
  fjournal = {Journal of Intelligent & Robotic Systems},
}

@inproceedings{estivill2018continuous,
  title        = {Continuous integration for testing full robotic behaviours in a GUI-stripped simulation},
  author       = {Estivill-Castro, V. and Hexel, R. and Lusty, C.},
  year         = 2018,
  booktitle    = models
}

@inproceedings{garcia2018icsa,
  title        = {An Architecture for Decentralized, Collaborative, and Autonomous Robots},
  author       = {Garcia, Sergio and Menghi, Claudio and Pelliccione, Patrizio and Berger, Thorsten and Wohlrab, Rebekka},
  year         = 2018,
  booktitle    = icsa
}

@Article{garousi2018testing,
  author  = {Garousi, V. and Felderer, M. and Karap{\i}{\c{c}}ak, C. M. and Y{\i}lmaz, U.},
  journal = ist,
  title   = {Testing embedded software: A survey of the literature},
  year    = {2018},
  volume  = {104},
}

@inproceedings{khosrowjerdi2018learning,
  title        = {Learning-based testing for autonomous systems using spatial and temporal requirements},
  author       = {Khosrowjerdi, Hojat and Meinke, Karl},
  year         = 2018,
  booktitle    = mases
}

@inproceedings{marcosig2018devs,
  title        = {Devs-over-ros (Dover): A Framework For Simulation-driven Embedded Control Of Robotic Systems Based On Model Continuity},
  author       = {Marcosig, Ezequiel Pecker and Giribet, Juan I. and Castro, Rodrigo},
  year         = 2018,
  booktitle    = wsc
}

@inproceedings{mullins2018accelerated,
  title        = {Accelerated Testing and Evaluation of Autonomous Vehicles via Imitation Learning},
  author       = {Mullins, Galen E. and Dress, Austin G. and Stankiewicz, Paul G. and Appler, Jordan D. and Gupta, Satyandra K.},
  year         = 2018,
  booktitle    = icra,
}

@inproceedings{santos2018esec,
  title        = {Property-Based Testing for the Robot Operating System},
  author       = {Santos, Andr{\'e} and Cunha, Alcino and Macedo, Nuno},
  year         = 2018,
  booktitle    = a-test
}

@inproceedings{timperley2018icst,
  title        = {Crashing {{Simulated Planes}} Is {{Cheap}}: {{Can Simulation Detect Robotics Bugs Early}}?},
  author       = {Timperley, Christopher Steven and Afzal, Afsoon and Katz, Deborah S. and Hernandez, Jam Marcos and Le Goues, Claire},
  year         = 2018,
  booktitle    = icst
}

@inproceedings{wienke2018irc,
  title        = {Model-{{Based Performance Testing}} for {{Robotics Software Components}}},
  author       = {Wienke, Johannes and Wigand, Dennis and Koster, Norman and Wrede, Sebastian},
  year         = 2018,
  booktitle    = irc
}

@Article{gucunski2017rabit,
  author  = {Gucunski, Nenad and Basily, Basily and Kim, Jinyoung and Yi, Jingang and Duong, Trung and Dinh, Kien and Kee, Seong-Hoon and Maher, Ali},
  journal = ijra,
  title   = {RABIT: Implementation, performance validation and integration with other robotic platforms for improved management of bridge decks},
  year    = {2017},
  number  = {3},
  volume  = {1},
}

@inproceedings{wei2017evolving,
  title        = {Evolving test environments to identify faults in swarm robotics algorithms},
  author       = {Hao Wei and Timmis, Jon and Alexander, Rob},
  year         = 2017,
  booktitle    = cec
}

@incollection{honfi2017sdl,
  title        = {Model-{{Based Regression Testing}} of {{Autonomous Robots}}},
  author       = {Honfi, D{\'a}vid and Moln{\'a}r, G{\'a}bor and Micskei, Zolt{\'a}n and Majzik, Istv{\'a}n},
  year         = 2017,
  booktitle    = sdl
}

@Article{kuehn2017system,
  author  = {Kuehn, Daniel and Schilling, Moritz and Stark, Tobias and Zenzes, Martin and Kirchner, Frank},
  journal = jfr,
  title   = {System Design and Testing of the Hominid Robot Charlie},
  year    = {2017},
  number  = {4},
  volume  = {34},
}

@InProceedings{mauch2017service,
  author    = {Mauch, Felix and Roennau, Arne and Heppner, Georg and Buettner, Timothee and Dillmann, R{\"{u}}diger},
  booktitle = icar,
  title     = {Service robots in the field: The BratWurst Bot},
  year      = {2017},
}

@inproceedings{moon2017usability,
  title        = {Usability evaluation of movement support service robot for elderly},
  author       = {Moon, Myung Kug and Kim, Seon-Chil},
  year         = 2017,
  booktitle    = aemusp
}

@inproceedings{sotiropoulos2017qrs,
  title        = {Can {{Robot Navigation Bugs Be Found}} in {{Simulation}}? {{An Exploratory Study}}},
  author       = {Sotiropoulos, Thierry and Waeselynck, Helene and Guiochet, Jeremie and Ingrand, Felix},
  year         = 2017,
  booktitle    = qrs
}

@Article{wienke2017ar,
  author  = {Wienke, J. and Wrede, S.},
  journal = adv-robotics,
  title   = {Performance Regression Testing and Run-Time Verification of Components in Robotics Systems},
  year    = {2017},
  number  = {22},
  volume  = {31},
}

@incollection{amigoni2016explorative,
  title        = {Explorative experiments in autonomous robotics},
  author       = {Amigoni, F. and Schiaffonati, V.},
  year         = 2016,
  booktitle    = mbr,
}

@inproceedings{araiza2016intelligent,
  title        = {Intelligent Agent-Based Stimulation for Testing Robotic Software in Human-Robot Interactions},
  author       = {Araiza-Illan, Dejanira and Pipe, Anthony G. and Eder, Kerstin},
  year         = 2016,
  booktitle    = morse,
}

@inproceedings{araiza2016systematic,
  title        = {Systematic and realistic testing in simulation of control code for robots in collaborative human-robot interactions},
  author       = {Araiza-Illan, Dejanira and Western, David and Pipe, Anthony G and Eder, Kerstin},
  year         = 2016,
  booktitle    = taros
}

@Article{ball2016vision,
  author   = {Ball, David and Upcroft, Ben and Wyeth, Gordon and Corke, Peter and English, Andrew and Ross, Patrick and Patten, Tim and Fitch, Robert and Sukkarieh, Salah and Bate, Andrew},
  journal  = {J. Field Robot.},
  title    = {Vision-based Obstacle Detection and Navigation for an Agricultural Robot},
  year     = {2016},
  number   = {8},
  volume   = {33},
  fjournal = {Journal of Field Robotics},
}

@inproceedings{maruyama2016emsoft,
  title        = {Exploring the Performance of {{ROS2}}},
  author       = {Maruyama, Yuya and Kato, Shinpei and Azumi, Takuya},
  year         = 2016,
  booktitle    = emsoft
}

@inproceedings{saglietti2016model,
  title        = {Model-driven Structural and Statistical Testing of Robot Cooperation and Reconfiguration},
  author       = {Saglietti, Francesca and Meitner, Matthias},
  year         = 2016,
  booktitle    = morse
}

@inproceedings{sotiropoulos2016virtual,
  title        = {Virtual Worlds for Testing Robot Navigation: A Study on the Difficulty Level},
  author       = {Sotiropoulos, Thierry and Guiochet, Guiochet and Ingrand, Ingrand and Waeselynck, Weaselynck},
  year         = 2016,
  booktitle    = edcc
}

@inproceedings{andrews2015active,
  title        = {Active World Model for Testing Autonomous Systems Using CEFSM},
  author       = {Andrews, A. and Abdelgawad, M. and Gario, A.},
  year         = 2015,
  booktitle    = modevva,

}

@Article{barr2014oracle,
  author  = {Barr, E. T. and Harman, M. and McMinn, P. and Shahbaz, M. and Yoo, S.},
  journal = {TSE},
  title   = {The Oracle Problem in Software Testing: A Survey},
  year    = {2015},
  number  = {5},
  volume  = {41},
}

@inproceedings{ernits2015emcr,
  title        = {Model-Based Integration Testing of ROS Packages: A Mobile Robot Case Study},
  author       = {Ernits, J. and Halling, E. and Kanter, G. and Vain, J.},
  year         = 2015,
  booktitle    = ecmr,
}

@inproceedings{paikan2015generic,
  title        = {A generic testing framework for test driven development of robotic systems},
  author       = {Paikan, Ali and Traversaro, Silvio and Nori, Francesco and Natale, Lorenzo},
  year         = 2015,
  booktitle    = mesas
}

@book{bath2014,
  title        = {The Software Test Engineer's Handbook: {{A}} Study Guide for the {{ISTQB}} Test Analyst and Technical Test Analyst Advanced Level Certificates},
  author       = {Bath, Graham and McKay, Judy},
  year         = 2014,
  publisher    = {Rocky Nook Inc.}
}

@incollection{bihlmaier2014simpar,
  title        = {Robot {{Unit Testing}}},
  author       = {Bihlmaier, Andreas and W{\"o}rn, Heinz},
  year         = 2014,
  booktitle    = simpar
}

@Article{fluckiger2014service,
  author   = {Fl{\"{u}}ckiger, Lorenzo and Utz, Hans},
  journal  = {J. Field Robot.},
  title    = {Service Oriented Robotic Architecture for Space Robotics: Design, Testing, and Lessons Learned},
  year     = {2014},
  number   = {1},
  volume   = {31},
  fjournal = {Journal of Field Robotics},
}

@book{hass2014guide,
  title        = {Guide to Advanced Software Testing},
  author       = {Hass, Anne Mette},
  year         = 2014,
  publisher    = {Artech House, Inc},
  isbn         = {1608078043}
}

@Article{jochmann2014virtual,
  author  = {Jochmann, G. and Bl{\"u}mel, F. and Stern, O. and Ro{\ss}mann, J.},
  journal = ki,
  title   = {The virtual space robotics testbed: Comprehensive means for the development and evaluation of components for robotic exploration missions},
  year    = {2014},
  number  = {2},
  volume  = {28},
}

@inproceedings{lill2014testing,
  title        = {Testing the cooperation of autonomous robotic agents},
  author       = {Lill, Raimar and Saglietti, Francesca},
  year         = 2014,
  booktitle    = icsoftea
}

@incollection{lochau2014sfm,
  title        = {Model-Based Testing},
  author       = {Lochau, Malte and Peldszus, Sven and Kowal, Matthias and Schaefer, Ina},
  year         = 2014,
  booktitle    = sfm
}

@inproceedings{sweet2014demonstration,
  title        = {Demonstration of prognostics-enabled decision making algorithms on a hardware mobile robot test platform},
  author       = {Sweet, Adam and Gorospe, George and Daigle, Matthew and Balaban, Edward and Roychoudhury, Indranil and Narasimhan, Sriram and others},
  year         = 2014,
  booktitle    = phmconf
}

@TechReport{iso29119pt1,
  author = {{ISO/IEC/IEEE}},
  title  = {{Software and Systems Engineering {\textendash}{{Software}} Testing -- Part 1: Concepts and Definitions}},
  year   = {2013},
  number = {ISO/IEC/IEEE 29119-1},
  type   = {Int. Std.},
}

@TechReport{iso29119pt2,
  author = {{ISO/IEC/IEEE}},
  title  = {{Software and Systems Engineering {\textendash}{{Software}} Testing -- Part 2: Test Processes}},
  year   = {2013},
  number = {ISO/IEC/IEEE 29119-2},
  type   = {Int. Std.},
}

@TechReport{iso29119pt3,
  author = {{ISO/IEC/IEEE}},
  title  = {{Software and Systems Engineering -- Software Testing -- Part 3: Test Documentation}},
  year   = {2013},
  number = {ISO/IEC/IEEE 29119-3},
  type   = {Int. Std.},
}

@Article{jimenez2013testbeds,
  author  = {Adri{\'{a}}n Jim{\'{e}}nez-Gonz{\'{a}}lez and Jose Ramiro {Martinez-de Dios} and Anibal Ollero},
  journal = ras,
  title   = {Testbeds for ubiquitous robotics: A survey},
  year    = {2013},
  number  = {12},
  volume  = {61},
}

@inproceedings{biggs2013experiences,
  title        = {Experiences with model-centred design methods and tools in safe robotics},
  author       = {Biggs, G. and Sakamoto, T. and Fujiwara, K. and Anada, K.},
  year         = 2013,
  booktitle    = iros
}

@Article{broggi2013extensive,
  author  = {Broggi, Alberto and Buzzoni, Michele and Debattisti, Stefano and Grisleri, Paolo and Laghi, Maria Chiara and Medici, Paolo and Versari, Pietro},
  journal = {T-ITS},
  title   = {Extensive Tests of Autonomous Driving Technologies},
  year    = {2013},
  number  = {3},
  volume  = {14},
}

@Article{poncela2013framework,
  author  = {Poncela, Javier and Aguayo-Torres, MC},
  journal = wpc,
  title   = {A framework for testing of wireless underwater robots},
  year    = {2013},
  number  = {3},
  volume  = {70},
}

@inproceedings{bostelman2013development,
  title        = {Development of standard test methods for unmanned and manned industrial vehicles used near humans},
  author       = {Roger Bostelman and Richard Norcross and Joe Falco and Jeremy Marvel},
  year         = 2013,
  booktitle    = mmif
}

@inproceedings{zendel2013vitro,
  title        = {VITRO - Model based vision testing for robustness},
  author       = {Zendel, Oliver and Herzner, Wolfgang and Murschitz, Markus},
  year         = 2013,
  booktitle    = isr
}

@Article{zenzeri2013using,
  author  = {Zenzeri, Jacopo and De Santis, Dalia and Mohan, Vishwanathan and Casadio, Maura and Morasso, Pietro},
  journal = jrobotics,
  title   = {Using the Functional Reach Test for Probing the Static Stability of Bipedal Standing in Humanoid Robots Based on the Passive Motion Paradigm},
  year    = {2013},
  number  = {1},
  volume  = {2013},
}

@inproceedings{bostelman2012standard,
  title        = {Standard test procedures and metrics development for automated guided vehicle safety standards},
  author       = {Bostelman, Roger and Shackleford, Will and Cheok, Geraldine and Norcross, Richard},
  year         = 2012,
  booktitle    = permis
}

@InProceedings{cossellnovel,
  author  = {Cossell, Stephen},
  title   = {A Novel Approach to Automated Systems Engineering on a Multi-Agent Robotics Platform using Enterprise Configuration Testing Software},
  year    = {2012},
  journal = {ACRA},
}

@inproceedings{kang2012web,
  title        = {Web-based automated black-box testing framework for component based robot software},
  author       = {Kang, Jeong Seok and Park, Hong Seong},
  year         = 2012,
  booktitle    = ubicomp
}

@article{khan2012comparative,
  title        = {A comparative study of white box, black box and grey box testing techniques},
  author       = {Khan, Mohd Ehmer and Khan, Farmeena},
  year         = 2012,
  journal      = ijcasa,
  volume={3},
  number={6},

}

@incollection{micskei2012amsta,
  title        = {A {{Concept}} for {{Testing Robustness}} and {{Safety}} of the {{Context-Aware Behaviour}} of {{Autonomous Systems}}},
  author       = {Micskei, Zolt{\'a}n and Szatm{\'a}ri, Zolt{\'a}n and Ol{\'a}h, J{\'a}nos and Majzik, Istv{\'a}n},
  year         = 2012,
  booktitle    = kes-amsta
}

@inproceedings{powell2012testing,
  title        = {Testing the Input Timing Robustness of Real-Time Control Software for Autonomous Systems},
  author       = {Powell, David and Arlat, Jean and Chu, Hoang Nam and Ingrand, Felix and Killijian, Marc-Olivier},
  year         = 2012,
  booktitle    = edcc
}

@Article{sunspiral2012development,
  author  = {SunSpiral, Vytas and Wheeler, D.W. and Chavez-Clemente, Daniel and Mittman, David},
  journal = jfr,
  title   = {Development and field testing of the FootFall planning system for the ATHLETE robots},
  year    = {2012},
  number  = {3},
  volume  = {29},
}

@inproceedings{weiss2012multi,
  title        = {Multi-relationship evaluation design: modeling an automatic test plan generator},
  author       = {Weiss, Brian A. and Schmidt, Linda C.},
  year         = 2012,
  booktitle    = permis
}

@inproceedings{wilson2012uav,
  title        = {UAV rendezvous: From concept to flight test},
  author       = {Wilson, Daniel B and Goktogan, AH and Sukkarieh, Salah},
  year         = 2012,
  booktitle    = acra
}

@inproceedings{tang2011testbed,
  title        = {A testbed for real-time autonomous vehicle PHM and contingency management applications},
  author       = {Tang, Liang and Hettler, Eric and Zhang, Bin and DeCastro, Jonathan},
  year         = 2011,
  booktitle    = phmconf
}

@Article{barfoot2011field,
  author  = {T. Barfoot and P. Furgale and B. Stenning and P. Carle and L. Thomson and G. Osinski and M. Daly and N. Ghafoor},
  journal = ras,
  title   = {Field testing of a rover guidance, navigation, and control architecture to support a ground-ice prospecting mission to Mars},
  year    = {2011},
  number  = {6},
  volume  = {59},
}

@Article{wettergreen2010design,
  author  = {David Wettergreen and Scott Moreland and Krzysztof Skonieczny and Dominic Jonak and David Kohanbash and James Teza},
  journal = ijrr,
  title   = {Design and field experimentation of a prototype Lunar prospector},
  year    = {2010},
  number  = {12},
  volume  = {29},
}

@inproceedings{jacoff2010comprehensive,
  title        = {Comprehensive standard test suites for the performance evaluation of mobile robots},
  author       = {Jacoff, Adam and Huang, Hui-Min and Messina, Elena and Virts, Ann and Downs, Anthony},
  year         = 2010,
  booktitle    = permis
}

@Article{leonard2010coordinated,
  author  = {Leonard, N. E. and Paley, D. A. and Davis, R. E. and Fratantoni, D. M. and Lekien, F. and Zhang, F.},
  journal = jfr,
  title   = {Coordinated control of an underwater glider fleet in an adaptive ocean sampling field experiment in Monterey Bay},
  year    = {2010},
  number  = {6},
  volume  = {27},
}

@Article{lim2010ijsea,
  author  = {Lim, Jae-Hee and Song, Suk-Hoon and Son, Jung-Rye and Kuc, Tae-Yong and Park, Hong-Seong and Kim, Hong-Seok},
  journal = ijsea,
  title   = {An Automated Test Method for Robot Platform and Its Components},
  year    = {2010},
  number  = {3},
  volume  = {4},
}

@Article{michael2010grasp,
  author  = {Michael, Nathan and Mellinger, Daniel and Lindsey, Quentin and Kumar, Vijay},
  journal = ram,
  title   = {The GRASP Multiple Micro-UAV Testbed},
  year    = {2010},
  number  = {3},
  volume  = {17},
}

@inproceedings{paul2010robotic,
  title        = {A robotic system for steel bridge maintenance: Field testing},
  author       = {Paul, Gavin and Webb, Stephen and Liu, Dikai and Dissanayake, Gamini},
  year         = 2010,
  booktitle    = acra
}

@incollection{piel2010automating,
  title        = {Automating integration testing of large-scale publish/subscribe systems},
  author       = {Piel, {\'E}ric and Gonz{\'a}lez, Alberto and Gross, Hans-Gerhard},
  year         = 2010,
  booktitle    = {Principles and Applications of Distributed Event-Based Systems}
}

@incollection{proetzsch2010lcns,
  title        = {A {{Systematic Testing Approach}} for {{Autonomous Mobile Robots Using Domain-Specific Languages}}},
  author       = {Proetzsch, Martin and Zimmermann, Fabian and Eschbach, Robert and Kloos, Johannes and Berns, Karsten},
  year         = 2010,
  booktitle    = ki
}

@Article{ruparelia2010software,
  author  = {Ruparelia, Nayan B},
  journal = acmsen,
  title   = {Software development lifecycle models},
  year    = {2010},
  number  = {3},
  volume  = {35},
}

@inproceedings{schlegel2010design,
  title        = {Design abstraction and processes in robotics: From code-driven to model-driven engineering},
  author       = {Schlegel, Christian and Steck, Andreas and Brugali, Davide and Knoll, Alois},
  year         = 2010,
  booktitle    = simpar
}

@inproceedings{ahuja2009test,
  title        = {Test results of autonomous behaviors for urban environment exploration},
  author       = {G. Ahuja and D. Fellars and G. Kogut and E. Pacis Rius and B. Sights and H. R. Everett},
  year         = 2009,
  booktitle    = spie-ust,
}

@Article{garcia2009implementation,
  author  = {Garcia, R.D. and Valavanis, K. P.},
  journal = jirs,
  title   = {The implementation of an autonomous helicopter testbed},
  year    = {2009},
  number  = {1},
  volume  = {54},
}

@Article{quigley2009icra,
  author  = {Quigley, Morgan and Gerkey, Brian and Conley, Ken and Faust, Josh and Foote, Tully and Leibs, Jeremy and Berger, Eric and Wheeler, Rob and Ng, Andrew},
  journal = icra-oss,
  title   = {{{ROS}}: An Open-Source {{Robot Operating System}}},
  year    = {2009},
  number  = {3.2},
  volume  = {3},
}

@inproceedings{schlegel2009robotic,
  title        = {Robotic software systems: From code-driven to model-driven designs},
  author       = {Schlegel, Christian and Ha{\ss}ler, Thomas and Lotz, Alex and Steck, Andreas},
  year         = 2009,
  booktitle    = icar
}

@Article{vitzilaios2009experimental,
  author  = {Vitzilaios, Nikos I and Tsourveloudis, Nikos C},
  journal = jirs,
  title   = {An experimental test bed for small unmanned helicopters},
  year    = {2009},
  number  = {5},
  volume  = {54},
}

@inproceedings{fink2009multi,
  title        = {Multi-rover testbed for teleconducted and autonomous surveillance, reconnaissance, and exploration},
  author       = {Wolfgang Fink and Mark A. Tarbell},
  year         = 2009,
  booktitle    = spie-set
}

@InProceedings{balakirsky2008integrated,
  author    = {Balakirsky, Stephen and Proctor, Frederick M. and Scrapper, Christopher J. and Kramer, Thomas R.},
  booktitle = idetc,
  title     = {An Integrated Control and Simulation Environment for Mobile Robot Software Development},
  year      = {2008},
}

@inproceedings{erdos2008uav,
  title        = {UAV Autopilot Integration and Testing},
  author       = {Erdos, David and Watkins, Steve E.},
  year         = 2008,
  booktitle    = ieee-r5
}

@Article{de2008distributed,
  author  = {Michael De Rosa and Seth Goldstein and Peter Lee and Jason Campbell and Padmanabhan Pillai},
  journal = ijrr,
  title   = {Distributed Watchpoints: Debugging Large Modular Robot Systems},
  year    = {2008},
  number  = {3-4},
  volume  = {27},
}

@conference{icinco08,
  title        = {Altitude Control Of Small Helicopters Using A Prototype Test Bed},
  author       = {Nikos I. Vitzilaios and Nikos C. Tsourveloudis},
  year         = 2008,
  booktitle    = icinco
}

@inproceedings{bertolino2007software,
  title        = {Software testing research: Achievements, challenges, dreams},
  author       = {Bertolino, Antonia},
  year         = 2007,
  booktitle    = fose
}

@inproceedings{gertman2007methodology,
  title        = {A methodology for testing unmanned vehicle behavior and autonomy},
  author       = {Gertman, D. I. and McFarland, C. and Klein, T. A. and Gertman, A. E. and Bruemmer, D. J.},
  year         = 2007,
  booktitle    = permis
}

@inproceedings{glinz2007non,
  title        = {On non-functional requirements},
  author       = {Glinz, Martin},
  year         = 2007,
  booktitle    = re
}

@inproceedings{humphrey2007assessing,
  title        = {Assessing the scalability of a multiple robot interface},
  author       = {Humphrey, Curtis M. and Henk, Christopher and Sewell, George and Williams, Brian W. and Adams, Julie A.},
  year         = 2007,
  booktitle    = hri
}

@Article{yu2007development,
  author  = {Yu, Seung Nam and Lee, Seung Yeol and Han, Chang Soo and Lee, Kye Young and Lee, Sang Heon},
  journal = auton-robots,
  title   = {Development of the curtain wall installation robot: Performance and efficiency tests at a construction site},
  year    = {2007},
  number  = {3},
  volume  = {22},
}

@inproceedings{kramer2006ugv,
  title        = {UGV acceptance testing},
  author       = {Kramer, Jeffrey A and Murphy, Robin R},
  year         = 2006,
  booktitle    = spie-ust
}

@Article{baresi2006introduction,
  author  = {Luciano Baresi and Mauro Pezz{\`{e}}},
  journal = {ENTCS},
  title   = {An Introduction to Software Testing},
  year    = {2006},
  number  = {1},
  volume  = {148},
}

@Article{metta2006yarp,
  author  = {Metta, Giorgio and Fitzpatrick, Paul and Natale, Lorenzo},
  journal = ijars,
  title   = {YARP: yet another robot platform},
  year    = {2006},
  number  = {1},
  volume  = {3},
}

@Article{montgomery2006jet,
  author  = {Montgomery, James F. and Johnson, Andrew E. and Roumeliotis, Stergios I. and Matthies, Larry H.},
  journal = jfr,
  title   = {The Jet Propulsion Laboratory Autonomous Helicopter Testbed: A platform for planetary exploration technology research and development},
  year    = {2006},
  number  = {3-4},
  volume  = {23},
}

@inproceedings{moshayedi2021simulation,
  title        = {Rapid Development of Vision-Based Control for MAVs through a Virtual Flight Testbed},
  author       = {Grzywna, J.W. and Jain, A. and Plew, J. and Nechyba, M.C.},
  year         = 2005,
  booktitle    = icra
}

@TechReport{kim2005rrt,
  title        = {An RRT-based algorithm for testing and validating multi-robot controllers},
  author       = {Kim, Jongwoo and Esposito, Joel M and Kumar, Vijay},
  year         = 2005,
  journal      = rssbook,
  
}

@Article{korayem2005vision,
  author  = {Korayem, M. H. and Khoshhal, K. and Aliakbarpour, H.},
  journal = ijamt,
  title   = {Vision based simulation and experiment for performance tests of robot},
  year    = {2005},
  number  = {11},
  volume  = {25},
}

@Article{bensalem2005testing,
  author  = {Saddek Bensalem and Marius Bozga and Moez Krichen and Stavros Tripakis},
  journal = {ENTCS},
  title   = {Testing Conformance of Real-Time Applications by Automatic Generation of Observers},
  year    = {2005},
  volume  = {113},
}

@inproceedings{jacoff2001reference,
  title        = {{Reference test courses for autonomous mobile robots}},
  author       = {Adam Jacoff and Elena Messina and John Evans},
  year         = 2001,
  booktitle    = spie-ugvt
}

@Article{mihali2000robotic,
  author  = {Mihali, Raul and Grigorian, Mher and Sobh, Tarek},
  journal = jirs,
  title   = {Robotic Optimization and Testing for the Formula One Tire-Changing Robot},
  year    = {2000},
  number  = {3},
  volume  = {29},
}

@Article{pook1999test,
  author  = {Polly K. Pook and Christopher K. DeBolt},
  journal = ijrr,
  title   = {Test Bed Robot Development for Cooperative Submunitions Clearance},
  year    = {1999},
  number  = {7},
  volume  = {18},
}

@Article{fink1997property,
  author  = {Fink, George and Bishop, Matt},
  journal = {Softw. Eng. Notes},
  title   = {Property-based testing: a new approach to testing for assurance},
  year    = {1997},
  number  = {4},
  volume  = {22},
}

@inproceedings{borenstein1995umbmark,
  title        = {{UMBmark: a benchmark test for measuring odometry errors in mobile robots}},
  author       = {Johann Borenstein and Liqiang Feng},
  year         = 1995,
  booktitle    = spie-mr
}

@Article{lamboley1995marsokhod,
  author  = {Lamboley, M and Proy, C and Rastel, L and Trong, T Nguyen and Zashchirinski, A and Buslaiev, S},
  journal = auton-robots,
  title   = {Marsokhod: Autonomous navigation tests on a Mars-like terrain},
  year    = {1995},
  number  = {4},
  volume  = {2},
}

@misc{istqb-glossary,
  title        = {Standard {{Glossary}} of {{Terms}} Used in {{Software Testing}}},
  author       = {{International Software Testing Qualifications Board (ISQTB)}},
  howpublished = {http://glossary.istqb.org/}
}

@manual{ros2testing,
  title        = {{Testing}},
  author       = {{ROS 2 Documentation}},
  note         = {Accessed: 2025-06-18},
  howpublished = {\url{https://docs.ros.org/en/jazzy/Tutorials/Intermediate/Testing/Testing-Main.html}}
}

@manual{rosunittest,
  title        = {{Automatic Testing with ROS}},
  author       = {{ROS Wiki}},
  note         = {Accessed: 2025-06-18},
  howpublished = {\url{http://wiki.ros.org/Quality/Tutorials/UnitTesting}}
}

@inproceedings{fornasier2021vinseval,
  title        = {VINSEval: Evaluation Framework for Unified Testing of Consistency and Robustness of Visual-Inertial Navigation System Algorithms},
  author       = {Fornasier, Alessandro and Scheiber, Martin and Hardt-Stremayr, Alexander and Jung, Roland and Weiss, Stephan},
  year         = {2021},
  booktitle    = icra, 
}

@Article{tang2020test,
  author   = {Tang, Shaomin and Liu, Guixiong and Lin, Zhiyu and Li, Xiaobing and Pan, Minqiang},
  journal  = {Sfi. S. Sci. C.},
  title    = {A Test Procedure Optimization Method for an Industrial Robot Servo System on an Integrated Testing Platform},
  year     = {2020},
  number   = {1},
  volume   = {2020},
  fjournal = {Complexity},
}

@Article{looije2017specifying,
  author  = {Rosemarijn Looije and Mark A. Neerincx and Koen V. Hindriks},
  journal = {Cogn. Syst. Res.},
  title   = {Specifying and testing the design rationale of social robots for behavior change in children},
  year    = {2017},
  volume  = {43},
}

@InProceedings{munoz2014first,
  author  = {Mu{\~{n}}oz, Pablo and Cesta, Amedeo and Orlandini, Andrea and R-Moreno, Mar{\'{i}}a D.},
  title   = {First Steps on an On-Ground Autonomy Test Environment},
  year    = {2014},
  journal = {SMC-IT/SCC},
}

@Article{mossige2015testing,
  author  = {Morten Mossige and Arnaud Gotlieb and Hein Meling},
  journal = ist,
  title   = {Testing robot controllers using constraint programming and continuous integration},
  year    = {2015},
  volume  = {57},
}

@Article{sartori2014implementation,
  author  = {Sartori, Daniele and Quagliotti, Fulvia and Rutherford, Matthew J and Valavanis, Kimon P},
  journal = JINT,
  title   = {Implementation and testing of a backstepping controller autopilot for fixed-wing UAVs},
  year    = {2014},
  number  = {3},
  volume  = {76},
}

@Article{straub2013characterization,
  author  = {Straub, Jeremy and Huber, Justin},
  journal = {Computers},
  title   = {A Characterization of the Utility of Using Artificial Intelligence to Test Two Artificial Intelligence Systems},
  year    = {2013},
  number  = {2},
  volume  = {2},
}

@Article{torta2012modeling,
  author  = {Torta, Elena and Cuijpers, Raymond H and Juola, James F and Van Der Pol, David},
  journal = IJHR,
  title   = {Modeling and testing proxemic behavior for humanoid robots},
  year    = {2012},
  number  = {04},
  volume  = {9},
}

@inproceedings{rodic2011scalable,
  title        = {Scalable experimental platform for research, development and testing of networked robotic systems in informationally structured environments experimental testbed station for wireless robot-sensor networks}, 
  author       = {Rodic, Aleksandar and Jovanovic, Milos and Popic, Svemir and Mester, Gyula},
  year         = {2011},
  journal      = RiiSS, 
}

@Article{martin2023verification,
  author  = {Enrique Martin-Martin and Manuel Montenegro and Adri{\'{a}}n Riesco and Juan Rodr{\'{i}}guez-Hortal{\'{a}} and Rub{\'{e}}n Rubio},
  journal = jlamp,
  title   = {Verification of the ROS NavFn planner using executable specification languages},
  year    = {2023},
  volume  = {132},
}

@Article{marquez2023hardware,
  author  = {Marquez, J. and Sullivan, C. and Price, R. M. and Roberts, R. C.},
  journal = ral,
  title   = {Hardware-in-the-Loop Soft Robotic Testing Framework Using an Actor-Critic Deep Reinforcement Learning Algorithm},
  year    = {2023},
  number  = {9},
  volume  = {8},
}

@Article{sibilska2022framework,
  author  = {Sibilska-Mroziewicz, Anna and Mo{\.z}aryn, Jakub and Hameed, Ayesha and Fern{\'a}ndez, Mar{\'\i}a Molina and Ordys, Andrzej},
  journal = mbd,
  title   = {Framework for simulation-based control design evaluation for a snake robot as an example of a multibody robotic system},
  year    = {2022},
  number  = {4},
  volume  = {55},
}

@InProceedings{carlone2009comparative,
  author    = {Carlone, Luca and Bona, Basilio},
  booktitle = {ICAR},
  title     = {A comparative study on robust localization: Fault tolerance and robustness test on probabilistic filters for range-based positioning},
  year      = {2009},
}

@InProceedings{totev2021testing,
  author    = {Nikola Totev},
  booktitle = {ISGT},
  title     = {Testing Robotic Systems: Case Study},
  year      = {2021},
  series    = {CEUR Workshop Proc.},
  volume    = {2933},
}

@inproceedings{allozi2022feasibility,
  title        = {Feasibility Analysis of Path Planning Algorithms}, 
  author       = {Allozi, Elbera and Yilmaz, Abdurrahman and Ervan, Osman and Temeltas, Hakan},
  booktitle    = INISTA, 
  year         = {2022},
}

@inproceedings{nybacka2009opportunities,
    author     = {Nybacka, Mikael},
    title      = {Opportunities in Automotive Winter Testing},
    booktitle  = IDETC/CIE,
    year       = {2009},
}

@Article{macenski2023regulated,
  author   = {Macenski, Steve and Singh, Shrijit and Mart{\'{i}}n, Francisco and Gin{\'{e}}s Clavero, Jonatan},
  journal  = {Auton. Robot.},
  title    = {Regulated Pure Pursuit for Robot Path Tracking},
  year     = {2023},
  number   = {6},
  volume   = {47},
  fjournal = {Autonomous Robots},
}

@Article{aljalbout2025reality,
  author  = {Aljalbout, E. and Xing, J. and Romero, A. and Akinola, I. and Garrett, C. R. and Heiden, E. and Gupta, A. and others},
  journal = {Annu. Rev. Control Robot. Auton. Syst.},
  title   = {The Reality Gap in Robotics: Challenges, Solutions, and Best Practices},
  year    = {2025},
}

@book{mcconnell1998software,
  title={Software project survival guide},
  author={McConnell, Steve},
  year={1998},
  publisher={Pearson Education}
}

@Article{sven2025reconfiguration,
  author  = {Sven Peldszus and Davide Brugali and Daniel Strueber and Patrizio Pelliccione and Thorsten Berger},
  journal = {EMSE},
  title   = {Software Reconfiguration in Robotics},
  year    = {2025},
  number  = {3},
  volume  = {30},
}

@Article{garcia.ea:2023:robotvar,
  author  = {Sergio Garcia and Daniel Strueber and Davide Brugali and Alessandro Di Fava and Patrizio Pelliccione and Thorsten Berger},
  journal = {EMSE},
  title   = {Software Variability in Service Robotics},
  year    = {2023},
  number  = {1},
  volume  = {28},
}

@Article{caldas2024runtime,
  author  = {Ricardo Caldas and Juan Antonio Pi{\~{n}}era Garc{\'{i}}a and Matei Schiopu and Patrizio Pelliccione and Gena{\'{i}}na Rodrigues and Thorsten Berger},
  journal = {TSE},
  title   = {Runtime Verification and Field Testing for ROS-Based Robotic Systems},
  year    = {2024},
  number  = {10},
  volume  = {50},
}

@incollection{garcia2019robotics,
  author = {Sergio Garcia and Daniel Strueber and Davide Brugali and Alessandro Di Fava and Philipp Schillinger and Patrizio Pelliccione and Thorsten Berger},
  booktitle = {VaMoS},
  title = {Variability Modeling of Service Robots: Experiences and Challenges},
  year = {2019}
}

@Article{Brugali2006Stability,
  author  = {Brugali, D. and Salvaneschi, P.},
  journal = {IJARS},
  title   = {Stable aspects in robot software development},
  year    = {2006},
  number  = {1},
  volume  = {3},
}

@InProceedings{educon2014SPL,
  author    = {Brugali, D.},
  booktitle = {EDUCON},
  title     = {Exploiting the synergies between Robotics and Software Engineering: a project-based laboratory},
  year      = {2014},
}

@book{brugali2007software,
  title={Software engineering for experimental robotics},
  author={Brugali, Davide},
  volume={30},
  year={2007},
  publisher={Springer}
}

@Article{miller1990empirical,
  author  = {Miller, Barton P. and Fredriksen, Lars and So, Bryan},
  journal = {Commun. ACM},
  title   = {An empirical study of the reliability of UNIX utilities},
  year    = {1990},
  number  = {12},
  volume  = {33},
}

@Article{queiroz2024driver,
  author  = {Queiroz, Rodrigo and Sharma, Divit and Caldas, Ricardo and Czarnecki, Krzysztof and Garc{\'{i}}a, Sergio and Berger, Thorsten and Pelliccione, Patrizio},
  journal = {T-ITS},
  title   = {A Driver-Vehicle Model for ADS Scenario-Based Testing},
  year    = {2024},
  number  = {8},
  volume  = {25},
}

@Article{kang2015automatic,
  author  = {Jeong Seok Kang and Hong Seong Park},
  journal = ist,
  title   = {Automatic Generation Algorithm of Expected Results for Testing of Component-Based Software System},
  year    = {2015},
  volume  = {57},
}

@TechReport{ieee61012,
  author = {{IEEE}},
  title  = {{Standard Glossary of Software Engineering Terminology}},
  year   = {1990},
  number = {IEEE Std 610.12-1990},
  type   = {Int. Std.},
}

\appendix
\newpage

\end{document}